\documentclass{article}
\usepackage{graphicx}
\usepackage{amsmath}
\usepackage{epsfig}
\usepackage{changebar}
\usepackage{amsfonts}
\usepackage{amssymb}

\begin{document}

\title{Configurational Entropy and its Crisis in Metastable States: Ideal Glass
Transition in a Dimer Model as a Paragidm of a Molecular Glass}
\author{F. Semerianov and P. D. Gujrati\\Department of Physics, Department of Polymer Science,\\The University of Akron, Akron, OH 44311, USA}
\date{\today}
\maketitle
\begin{abstract}
We discuss the need for discretization to evaluate the configurational entropy
in a general model. We also discuss the prescription using restricted
partition function formalism to study the stationary limit of metastable
states. We introduce a lattice model of dimers as a paradigm of molecular
fluid and study metastability in it to investigate the root cause of glassy
behavior. We demonstrate the existence of the entropy crisis in metastable
states, from which it follows that the entropy crisis is the root cause
underlying the ideal glass transition in systems with particles of all sizes.
The orientational interactions in the model control the nature of the
liquid-liquid transition observed in recent years in molecular glasses.
\end{abstract}

\section{Introduction}

Glass transition (GT) in a glass-forming system such as a single-component
liquid (for example, water and silicate melts) remains a controversial
long-standing problem, even after many decades of active investigation and
presents one of the most challenging problems in theoretical physics[1-4]. In
particular, the existence of high and low density forms of viscous water [5]
is a consequence of a liquid-liquid transition in the glassy state [6]. As the
liquid is cooled below its melting temperature $T_{\text{M}}$ with sufficient
care so that the crystallization does not occur, the liquid gets in a
metastable liquid (ML) state, commonly known as the supercooled liquid (SCL)
state, which is \emph{disordered} with respect to its \emph{ordered}
crystalline phase (CR). The glass transition occurs in SCL at a temperature
\textit{T}$_{\mathrm{G}},$ which is usually about two-thirds of the melting
temperature $T_{\text{M}}$ for the liquid. The relaxation time and the
viscosity increase by several orders of magnitudes, typically within a range
of a few decades of the temperature as it is lowered, and eventually surpass
experimental limits. In other words, the system basically freezes at
\textit{T}$_{\mathrm{G}}$ \emph{without} any anomalous changes in its
thermodynamic densities like its specific volume or the entropy density. In
particular, no spatial correlation length has been identified so far that
would diverge near the transition at \textit{T}$_{\mathrm{G}}$. On the other
hand, mode-coupling theory [7] shows a dynamic slowdown at a temperature
\textit{T}$_{\mathrm{MC}}$ much higher than \textit{T}$_{\mathrm{G}}$ but
lower than $T_{\text{M}}$, and a loss of ergodicity. The relationship between
\textit{T}$_{\mathrm{MC}}$ and \textit{T}$_{\mathrm{G}}$ is not understood at
present, although attempts have been made recently [8-10] to understand it
partially in connection with long polymers. The free volume falls rapidly near
\textit{T}$_{\mathrm{MC}}$ for long polymer fluids, with the nature of the
drop becoming singular in the limit of infinitely long polymers. The glass
transition occurs at a temperature $T_{\text{K}}<$\textit{T}$_{\mathrm{MC}}$
where the configurational entropy vanishes, even if the polymer liquid is
incompressible [8] at all temperatures. The configurational entropy $S(T),$ by
definition [3], is the entropy that is used to define the
\emph{configurational partition function} (PF) \
\begin{equation}
Z(T)\equiv\frac{1}{V^{N}}\int^{^{\prime}}e^{-\beta E}d^{N}\{\mathbf{r\},}
\label{CPF}%
\end{equation}
in the canonical ensemble for a system of $N$ particles in a volume $V$ at a
given temperature $T$. The prime indicates that the integration is over
distinct configurations, $E(\{\mathbf{r}_{i}\})$ represents the potential
energy in a given configuration $\{\mathbf{r}_{i}\}$ determined by the
instantaneous positions of all the particles, $d^{N}\{\mathbf{r\}}$ represent
integrations with respect to $N$ positions $\mathbf{r}_{i}$ of the particles,
and $\beta\equiv1/T$, $T$ being the inverse temperature in the units of the
Boltzmann constant $k_{\text{B}}.$ The number of distinct configurations
$W(E)dE$ with energy in the range $E$ and $E+dE$ determines the
configurational entropy $S(E)\equiv\ln W(E)dE\simeq\ln W(E)$ [11] in the
corresponding microcanonical ensemble at a given configurational energy $E.$
For the state to be realizable in Nature, it is obvious that the number of
states must be a positive integer. Hence, we must always have $S(E)\geq0.$ The
average energy $\overline{E}(T)$ in the canonical ensemble gives the
configurational entropy $S(T)\equiv S[\overline{E}(T)]$ in that ensemble, and
must also be non-negative$.$

The loss of configurational entropy seems to be a generic phenomenon. The
thermal data for various systems capable to form glassy states exhibit an
\emph{entropy crisis} discovered by Kauzmann[1], in which a rapid entropy drop
to a \emph{negative} value occurs below the glass transition temperature in
SCL [2,4]. This strongly suggesting a deep connection between thermodynamics
and GT, since thermodynamics requires the entropy to decrease as the
temperature is lowered. Consequently, it is tempting to treat the
experimentally observed GT as a manifestation of an underlying thermodynamic
\emph{ideal glass transition}, which is invoked to avoid the genuine entropy
crisis when the SCL \emph{configurational entropy }$S_{\text{SCL}}<0$ below a
non-zero temperature $T=T_{\text{K}}.$ Negative $S_{\text{SCL}}$ implies that
such states are \emph{unrealizable} in Nature [8-10,12-13], which is what the
entropy crisis signifies. Free volume, which also seems to decrease with
lowering the temperature [1-2,4], does not seem to be the primary cause,
though it may be secondary, behind GT as shown recently [10], as there is no
thermodynamic requirement for it to decrease with the lowering of the temperature.

The entropy crisis and the resulting ideal GT have \emph{only }been
substantiated so far in two exact calculations. The first one is an exact
calculation by Derrida on an abstract model known as the random energy model
[14]. The second one is an explicit exact calculation by Gujrati and coworkers
[8-10] in long semiflexible polymers, which not only satisfies the rigorous
Gujrati-Goldstein free energy upper bound [15] for the equilibrium state, but
also yields $S_{\text{SCL}}(T)<0$ below a positive temperature $T=T_{\text{K}%
}$ [13]. An earlier demonstration of the crisis in polymers by Gibbs and
DiMarzio [16] was severely criticized by Gujrati and Goldstein [15] for its
poor approximation that violates the rigorous Gujrati-Goldstein bounds, thus
raising doubts about their primary conclusion of demonstrating an entropy
crisis [13]. Despite its limitation, the work has played a pivotal role in
elevating the Kauzmann entropy crisis from a mere curious observation to
probably the most important mechanism behind the glass transition, even though
the demonstration was only for long molecules. Unfortunately, the criticism by
Gujrati and Goldstein has been incorrectly interpreted by some workers [17] by
taking their bounds to be also applicable to metastable states in polymers. To
overcome the bounds, DiMarzio and Yang have suggested to replace the
unrealizability condition $S_{\text{SCL}}<0$ by an arbitrary condition
$S_{\text{SCL}}<S_{\text{c0}}$ for the entropy crisis, where $S_{\text{c0}}$
is some small critical value [13]. This is not the correct interpretation of
the Gujrati-Goldstein bounds. These bounds are only for the equilibrium
states, since they are obtained by considering Gujrati-Goldstein excitations
in CR; they are not applicable to SCL, which represents a disordered state.

A glass can be thought as a disordered or amorphous solid [18]. There are no
general arguments [18-19] to show that thermodynamically stable states must
always be ordered, i.e., periodic. The remarkable aperiodic Penrose
tilings\ of the plane, for example, by two differently but suitably shaped
tiles are stable. Nevertheless, it appears that most amorphous states in
systems that crystallize have much \emph{higher} internal energy or the
enthalpy than the corresponding crystal, even near or at absolute zero [1,18].
Thus, we will consider amorphous solids, i.e. glasses, as representing
metastable, rather than stable states in this work. The higher energy of the
glass has been used as the basis of a recent thermodynamic proof [12] that
there must exist a non-zero Kauzmann temperature $T_{\text{K}}$ below which
the configurational entropy of the disordered or the amorphous state of the
system becomes negative.

It should be stressed that the glass transition is ubiquitous and is also seen
in small molecules. Therefore, it is widely believed that the entropy crisis
also occurs in molecular liquids. However, no such entropy crisis has ever
been demonstrated in any explicit calculation for small molecules [4]. This is
highly disconcerting and casts doubts on the importance of the entropy crisis
for GT in systems consisting of particles of any size. It is this lack of
explicit demonstration of the entropy crisis in molecular liquids that has
motivated this work.

There is a clear need to settle whether the entropy crisis is generic or not
for molecules of all sizes. An explicit calculation without any uncontrollable
approximation such as an exact calculation for small molecules will go a long
way to settle the matter once for all. Without such a calculation, our
understanding of GT \emph{cannot} become complete. An exact calculation
demonstrating a positive Kauzmann temperature for small molecular fluids would
be a major accomplishment. The other motivation for the work is to carefully
define the configurational entropy so that is satisfies $S(T)\geq0$ for states
that are realizable in Nature$.$ Its \emph{violation} is then argued as a
trigger for the glass transition.

Our aim here is to fill this gap in our understanding by demonstrating the
existence of an entropy crisis in an explicit \emph{exact calculation} on a
system of classical dimers as a paradigm of a molecular liquid. In conjunction
with our earlier demonstration of the entropy crisis in infinitely and also
very long polymer system, our calculation, thus, finally enshrines the
Kauzmann entropy crisis ( negative $S_{\text{SCL}}$ implying that such states
are \emph{unrealizable} in Nature) as the general underlying thermodynamic
driving force for GT in molecules of all sizes. It should be stated that
recent computer simulations have not been able to settle the issue clearly
[20]. Our investigation also enables us to draw important conclusions about
the role orientational order plays in liquid-liquid (L-L) phase transitions
that have been observed recently in many atomic and molecular liquids [21-22].
It has gradually become apparent that the short-ranged orientational order in
supercooled liquids plays an important role not only in the formation of
glasses\ but also in giving rise to a liquid-liquid phase transition. Tanaka
[23] has proposed a general view in terms of cooperative medium-ranged bond
ordering to describe liquid-liquid transitions, based on the original work of
Nelson [24]. Thus, we will also consider orientational order in this work.

It is fair to say that there yet exists no completely satisfying theory of the
glass transition [3-4,25] even though some major progress has been made
recently [26-30,8-10,12]. Theoretical investigations mainly utilize two
different approaches, which are based either on thermodynamic or on kinetic
ideas. The two approaches provide an interesting duality in the liquid-glass
transition, neither of which seems complete. We discuss below briefly some of
the most promising theories based on these approaches.

\subsection{ Free Volume Theory}

The most successful theory that attempts to describe \textit{both} aspects
with some respectable success is based on the ''free-volume'' model of Cohen
and Turnbull [31]. The concept of free volume has been an intriguing one that
pervades throughout physics but its consequences and relevance are not well
understood [32], at least in our opinion, especially because there is no
consensus on what various workers mean by free volume. Nevertheless, GT in
this theory occurs when the free volume becomes sufficiently \textit{small} to
impede the mobility of the molecules [33]. The time-dependence of the
free-volume redistribution, determined by the energy barriers encountered
during redistribution, provides a \textit{kinetic} view of the transition, and
must be properly accounted for. This approach is yet to be completed to
satisfaction. Nevertheless, assuming that the change in free volume is
proportional to the difference in the temperature $T-T_{V}$ near the
temperature $T_{V},$ even though there is no thermodynamic requirement for the
free volume to drop as $T$\ is lowered [10], the viscosity $\eta(T)$ diverges
near $T_{V}$ according to the Vogel- Tammann-Fulcher equation%

\begin{equation}
\ln\eta(T)=A_{\text{VTF}}+B_{\text{VTF}}/(T-T_{V}), \label{eta1}%
\end{equation}
where $A_{\text{VTF}}$ and $B_{\text{VTF}}$ are system-dependent constants.
This situation should be contrasted with the fact that there are theoretical
models [14,8-9] \emph{without} any free volume in which the ideal glass
transition occurs due to the entropy crisis at a positive temperature. Hence,
it appears likely that the free volume itself is not the determining cause for
the glass transition in supercooled liquids. However, too much free volume can
destroy the transition [10]. We will, thus, explore the influence of free
volume on the glass transition in this work.

\subsection{Thermodynamic Theory}

The thermodynamic ideas alone describe GT in terms of the \emph{entropy crisis
}at a positive temperature [1]. The entropy crisis corresponds to a rapid
entropy drop to a \emph{negative} value below the glass transition temperature
in the supercooled liquid (SCL) [1-4]. An \emph{ideal glass transition} at
$T_{\text{K}}$ is invoked to avoid the entropy crisis. The entropic view plays
a central role in the Adams-Gibbs theory [34], according to which the
viscosity $\eta(T)$ above the glass transition depends on a quantity also
called the configurational entropy $S_{\text{conf}}(T)$ as follows:%

\begin{equation}
\ln\eta(T)=A_{\text{AG}}+B_{\text{AG}}/TS_{\text{conf}}(T), \label{eta2}%
\end{equation}
where $A_{\text{AG}}$ and $B_{\text{AG}}$ are system-dependent constants. The
existence of the entropy crisis has been justified in many exact model
calculations [14,8-10,16]. An alternative thermodynamic theory for the
impending entropy crisis based on spin-glass ideas has also been developed in
which proximity to an underlying first-order transition is used to explain the
glass transition [26].

In a thermodynamic theory, the glass-forming system is treated as homogeneous,
mainly because there appears to be no evidence of any structural changes
emerging near GT. From this point of view, a thermodynamic theory gives rise
to a homogeneous free energy. However, the fluctuations in it can be used to
construct the equations for time-dependence in the system. Thus, the
thermodynamic approach can be used to describe the dual aspect of the glass
transition. In particular, it should enable us to provide a bridge between the
two expressions in (2,3) for the viscosity that are the most widely-used and
successful formalism of viscosity in glass-forming liquids.

\subsection{Mode-Coupling Theory}

The mode-coupling theory [7] is an example of theories based on kinetic ideas
in which glass formation is described as a slowing down of liquid's mobility
and its freezing into a unique amorphous configuration at a positive
temperature. In this theory, the ergodicity is lost completely, and structural
arrest occurs at a temperature \textit{T}$_{\mathrm{MC}}$, which lies well
above the customary glass transition temperature \textit{T}$_{\mathrm{G}}$.
Consequently, the correlation time and the viscosity diverge due to the
\textit{caging effect}. The diverging viscosity can be related to the
vanishing free volume [31,33], which might suggest that \textit{the MC
transition is the same as the glass transition}. This does not seem to be the
consensus at present. Thus, it is not clear if the free volume is crucial for
the MC transition. Some progress has been made in this direction recently
[10], where it has been shown for long polymers that the free volume vanishes
at a temperature much higher than the ideal glass transition. The
mode-coupling theory is also not well-understood, especially below the glass
transition. More recently, it has been argued that this and mean-field
theories based on an underlying first-order transition may be incapable of
explaining dynamic heterogeneities.

\subsection{Theory vs Experiment/Simulation}

In this work, we are interested in the thermodynamic approach to investigate
the entropy crisis, according to which the configurational entropy
$S_{\text{SCL}}(T)$ vanishes at some positive temperature. It should be
remarked that the entropy crisis can only be seen in theoretical calculations,
but never in an experiment or simulation, since the latter two always deal
with states that are observed or produced, so that the corresponding entropy
would never be negative. It is only by extrapolation that the latter two may
predict an entropy crisis. The reliability of such extrapolation is debatable,
and has been used to argue against an entropy crisis [35-36]. By analyzing
experimental data, they have argued on procedural grounds that an entropy
crisis in any experiment must be absent, which we agree with. However, such
arguments based solely on experimental data or simulation \emph{without
extrapolation} can never shed light on the issue of the entropy crisis, which
is purely hypothetical. To verify the existence or non-existence of the
crisis, one must resort to theoretical arguments. Several workers [37-38] have
argued theoretically that it is not possible to have any entropy crisis, not
withstanding the explicit demonstration of it in long polymers, and in the
abstract random energy model. The argument due to Stillinger [37], in
particular, is forceful though not rigorous [4]. While he concedes that long
polymers may very well have an entropy crisis at a positive temperature, he
argues for its absence in viscous liquids of small molecules. From a purely
mathematical point of view, it is hard to understand how this scenario could
be possible. Using the physical argument of continuity, we expect
$T_{\text{K}}$ to be a smooth function of the molecular weight. Thus, it does
not seem possible that such a function remains zero over a wide but finite
range of the size of the molecules, and abruptly becomes non-zero for very
large sizes. A function like this must be a singular function. However, no
argument that we can imagine can support a particular large molecular size to
play the role of the location of such a singularity.

Assuming the entropy in (3) to be the configurational entropy, which is known
to vanish in theoretical calculations, then its rapid decrease to zero should
give rise to a diverging viscosity, thus providing a connection between the
thermodynamic view and \ the kinetic view. However, no experimental evidence
in support of such a diverging viscosity is known mainly because the
relaxation times become much longer than the experimental time limitations.
Thus, it does not seem surprising that theoretical predictions of entropy
crisis is never going to be seen directly in experiments.

Despite this, the suggestion that the rapid rise\ in the viscosity is due to a
sudden drop in $S_{\text{T}}(T)$ seems very enticing, since both phenomena are
ubiquitous in glassy states. The experimental data indicate that $T_{V}$\ and
$T_{K}$\ are, in fact, very close [39], clearly pointing to a close
relationship between the rapid rise in the viscosity and the entropy crisis.
This deep connection, if true, provides a very clean reflection of the dual
aspects of the glassy behavior mentioned above. It also implies strongly that
it is the entropy crisis, which is the root cause of the glass transition.
This idea got a strong support recently when it was shown that the free volume
could not be the root cause for GT in SCL, as the transition can exist even in
the absence of the free volume [10]. Thus, we are driven to the conclusion
that it should be possible to treat the SCL glass transition within a
thermodynamics formalism by demonstrating the existence of the entropy crisis.
In other words, if the scenario is valid, we can treat the experimentally
observed GT as a manifestation of the underlying ideal glass transition
induced by the entropy crisis. The ideal glass transition is a hypothetical
transition obtained in the limit of infinite slow cooling, provided the
crystal (CR) is forbidden to nucleate. This transition will never be observed
in experiments since such cooling rates are impossible to maintain in reality
or in simulations since any state generated in simulation will ensure that the
entropy is not negative. Thus, experimentally or in simulations, one would
never observe the ideal glass transition directly. It can only be inferred
either by some sort of extrapolation or by diverging relaxation times in these
methods. On the other hand, a theoretical demonstration of an ideal glass
transition is possible since neither the time restriction is an issue nor the
realizability of a state. The theoretical existence of an ideal glass
transition forces us to conclude that the observed glass transitions in
experiments are a manifestation of this transition, or in other words, of the
entropy crisis. As such, a study of the glass transition within a
thermodynamics formalism will enable us to understand glassy phenomena in a
systematic and fundamental way.

It should be noted that what one measures in experiments is the difference in
the entropy, and not the absolute entropy. Assuming that the entropy is zero
at absolute zero in accordance with the Nernst-Planck postulate, one can then
determine the absolute entropy experimentally. However, it is well known that
SCL is a metastable state, and there is no reason for its entropy to vanish at
absolute zero [18]. Indeed, it has been demonstrated some time ago that the
residual entropy at absolute zero obtained by extrapolation is a non-zero
fraction of the entropy of melting [40], which is not known a priori.
Therefore, it is \emph{impossible} to argue from experimental data that the
entropy indeed falls to zero, since such a demonstration will certainly
require calculating absolute entropy though efforts continue to this date [35-36].

The layout of the paper is the following. In the next section, we discuss
classical statistical mechanical approach to study glass transition and show
the need for discretizing both the real and the momentum spaces. We also
compare the configurational entropy in (1) and the criterion for the entropy
crisis with other definitions and criteria available in the literature. In
Section 3, we describe the classical dimer model, and its Husimi cactus
approximation is discussed in Section 4. We solve the model exactly on the
cactus by using the recursion relation (RR) technique and the results are
presented in Section 5. The final section contains discussion of our results
and conclusions. The RR's are deferred to the Appendix.

\section{Configurational Partition Function and Entropy Crisis}

\subsection{Negative Entropy in Continuum Classical Statistical Mechanics}

We consider a system of $N$ identical particles $i=1,2,...,N$ confined in a
container of volume $V$. The position $\mathbf{r}_{i}$ and momentum
$\mathbf{p}_{i} $ of the particle $i$ are treated as continuous variables in
classical mechanics. The Hamiltonian of the system is given by%

\begin{equation}
H\equiv\underset{i}{\sum}\mathbf{p}_{i}^{2}/2m+E(\{\mathbf{r}_{i}\}).
\label{H}%
\end{equation}
The first term in (4) represents the kinetic energy $K$. In classical
statistical mechanics (CSM), the total partition function (PF) $Z_{\text{T}}$
for the system can be reduced to a product of two different dimensionless
integrals as$\qquad\qquad\qquad\qquad\qquad\qquad\qquad$%
\begin{equation}
Z_{\text{T}}\equiv Z_{\text{KE}}Z\mathbf{,} \label{PF}%
\end{equation}
where%

\begin{equation}
Z_{\text{KE}}\equiv\frac{V^{N}}{(2\pi\hslash)^{3N}}\int e^{-\beta K}%
d^{N}\{\mathbf{p\}} \label{TRPF0}%
\end{equation}
denotes the PF due to the translational degrees of freedom in which
$d^{N}\{\mathbf{p\}}$ represents integrations with respect to $N$ momenta
$\mathbf{p}_{i}$ of the particles. The prefactor in terms of $\hslash$ is used
not only to make $Z_{N}$ dimensionless, but also to explicitly show the
correspondence of $Z_{N}$ with the corresponding PF in the quantum statistical
mechanics (QSM) in the classical limit $\hslash\rightarrow0$. Despite the
classical limit requirement $\hslash\rightarrow0$, we are not allowed to set
$\hslash=0$ in the final result, but keep its actual non-zero value.
Accordingly, some problems remain such as Wigner's distribution function not
being a classical probability distribution, which we do not discuss any
further but refer the reader to the literature [41]. Keeping $\hslash$ at its
non-zero value avoids infinities as we will see below but in no way implies
that we are dealing with quantum effects. In particular, it does not imply
that the entropy is non-negative, as can be seen in the following. The PF
$Z_{\text{KE}}$ can be written in terms of $W_{\text{KE}}(P)dP$ $\equiv
C_{3N}V^{N}P^{3N-1}dP/h^{3N},$ which is usually thought of as representing the
\emph{number of microstates} corresponding to the magnitude of the total
momentum $P$ in the range $P$, and $P+dP$ in the 3$N$-dimensional momentum
space, as follows:%

\begin{equation}
Z_{\text{KE}}\equiv\int W_{\text{KE}}(P)e^{-\beta K}dP; \label{TRPF}%
\end{equation}
here $C_{d}\equiv$ $d\pi^{d/2}/\Gamma(d/2+1)$ is a constant due to angular
integration in a $3N$-dimensional space. Being a number of states,
$W_{\text{KE}}(P)dP$ should be a \emph{positive} integer, which as we see
below is not true. The entropy function is given by $S_{\text{KE}}(P)=\ln
W_{\text{KE}}(P)$ [11], where we have used $K\equiv P^{2}/2m.$ In the
thermodynamic limit $N\rightarrow\infty$, the integrand in (7) must be
maximum. Hence, we look for the maximum of $P^{3N-1}e^{-\beta K}$ by the
prescription, known commonly as the saddle point approximation, of equating
its derivative to zero. This immediately gives the ideal gas equation for the
average kinetic energy $\overline{K}(T)$
\begin{equation}
\overline{K}(T)=(3/2)NT,
\end{equation}
which should come as no surprise. For the ideal gas, the total entropy
function $S_{\text{T}}(T)$ is given by%

\begin{equation}
S_{\text{T-ideal}}(T)=(3N/2)[\ln T+2V/3N+\ln(2\pi me/\sqrt{h})],
\label{IdealS}%
\end{equation}
while%

\begin{equation}
S_{\text{KE}}(T)=(3N/2)[1+\ln T+\ln(2\pi m/h^{2})]. \label{IdealSKE}%
\end{equation}
It should be noted that $S_{\text{KE}}(T)$ has the same value at a given
temperature for any classical system, regardless of the configurational
energy. Thus, in general, the configurational entropy can be always obtained
by subtracting $S_{\text{KE}}(T)$ from $S_{\text{T}}(T)$ [10]:%

\begin{equation}
S(T)\equiv S_{\text{T}}(T)-S_{\text{KE}}(T). \label{SCF}%
\end{equation}
For an ideal gas, the configurational PF is $Z=V^{N}/N!,$ and the
corresponding configurational entropy, which does not depend on $T$ now, is%

\begin{equation}
S_{\text{ideal}}=N\ln(Ve/N). \label{IdealSCF}%
\end{equation}
Both entropies are extensive, and so is the total ideal gas entropy
$S_{\text{T-ideal}}(T)$. If we set $h=0,$ we encounter an infinity in the
entropy at all temperatures in (10). To avoid this, we keep $h$ at its
non-zero value. At absolute zero, or for $V/N<1/e$, $S_{\text{T}}%
\rightarrow-\infty,$ a well-known result of classical statistical mechanics.
For $S_{\text{ideal}}$, it appears at first glance that the problem is due to
the point-like nature of the particles, which allows us to pack as many
particles as we wish in a given fixed volume $V$. This is not true. The
problem is due to the continuum nature of the real space.\ This is easily seen
from the exact solution of the 1-d Tonks gas, which is a simple model of
non-interacting hard rods, each of length $a$. In one dimension, the
configurational entropy $S$ corresponding to $N$ rods in a line segment of
length $L$ \ is given by [42]%

\begin{equation}
S_{\text{Tonks}}=N\ln[(L-Na)e/N] \label{TonksSCF}%
\end{equation}
in the thermodynamic limit. Comparison with (12) shows that the only
difference is that the total volume $V$ in (12) is replaced by the free volume
analog\ $L-Na$ in 1-dimensional Tonks gas (13). It is clear that the entropy
becomes \emph{negative} as soon as $L/N<1/e+a$ and eventually diverges to
$-\infty$ in the fully packed state. Similarly, the problem with
$S_{\text{KE}}(T)\rightarrow-\infty$ as $T\rightarrow0$ is due to the
continuum nature of the momentum space.

This problem disappears as soon as we invoke quantum statistical mechanics to
describe the total PF, which no longer can be written as product of two or
more PF's as in (5). Here, we consider the number of states $W_{\text{T}%
}(E_{\text{T}})\geq1$ as function of the (total) energy eigenvalue
$E_{\text{T}}.$ The energy eigenvalue $E_{\text{T}}$ can certainly be broken
into the kinetic energy part $K$ and the potential energy part $E$, but such a
partition is not possible for the total entropy $S_{\text{T}}(E_{\text{T}%
})\equiv\ln W_{\text{T}}(E_{\text{T}}).$ Therefore, in general, the notion of
the configurational entropy does not make sense in this case. Since we are
only concerned with classical statistical mechanics in this work, we will not
discuss this point further here.

\subsection{Non-negative Entropy in Classical Statistical Mechanics}

As shown above, the entropy of a continuum model is invariably negative in
classical statistical mechanics, especially at low temperatures; but this has
nothing to do with the entropy crisis noted by Kauzmann. Quantum statistical
mechanical calculations are not feasible at present to make entropy
non-negative. Thus, at present, no theoretical calculation using continuum
classical statistical will be able to justify the entropy crisis, as the
negative entropy may just be a manifestation of the continuum picture.
However, a lattice picture ensures a non-negative entropy for any state that
can be realized in Nature. Thus, a lattice model is capable of settling
whether an entropy crisis occurs in the metastable state or not.

To define the number of microstates so that it is always greater than or equal
to 1, one must \emph{discretize} the real and momentum spaces carefully. For
this purpose, we introduce a short distance $a$ in real space, and a related
quantity $b\equiv\hslash/2\pi a$ having the dimension of the momentum so that
$ab=2\pi\hslash$; $a$ may be taken as hard sphere of the molecule. We now
divide the two spaces into cells of size $a^{3N}$ and $b^{3N},$ respectively.
In the discretized space, the number of microstates is nothing but the number
of distinct ways the cells are occupied by the particles' positions and
momenta. This number \emph{cannot }be less than unity if the microstate is
realizable in Nature. Hence, the entropy can never be negative if the state
occurs in Nature. For example, $C_{3N}P^{3N-1}dP$ is now replaced by the
number of cells of size $b^{3N}$ that cover a shell of integer radius
$\overline{P}=P/b$, and thickness $b$ (corresponding to the integer thickness
$\overline{dP}=dP/b=1)$. This number is $C_{3N}\overline{P}^{3N-1}.$ The
equilibrium is obtained by maximizing the integrand
\[
I(\overline{P})\equiv\overline{P}^{3N-1}e^{-(\beta b^{2}/2m)\overline{P}^{2}%
}.
\]
Maximizing $I$ is equivalent to maximizing
\begin{equation}
\widetilde{I}\equiv I(\overline{P})/I(1)\equiv\overline{P}^{3N-1}e^{-(\beta
b^{2}/2m)(\overline{P}^{2}-1)}, \label{CorrectS}%
\end{equation}
in which $\overline{P}^{2}-1\geq0.$ It is evident that the maximum of
$\widetilde{I}$ at $\beta\rightarrow\infty$ corresponds to $\overline{P}=1,$
and gives $\widetilde{I}$ =1. This is true for all $N$, and will remain true
even as $N\rightarrow\infty$. In this limit, higher values of $\overline{P}$
also condense to the state corresponding to the lowest energy density per
particle. Thus, the number of microstates at absolute zero is strictly a
positive integer. Correspondingly, the entropy is\emph{\ not} negative
anymore. It is also clear that for small values of $\overline{P}$ (and finite
$N$), we cannot differentiate $\widetilde{I}$ with respect to $\overline{P}$
to find the maximum, as $\overline{P}$ cannot be approximated as a continuous
variable in this range. Thus, care must be exercised near absolute zero.

A similar conclusion is drawn for the configurational entropy determining $Z$
in (5). It is also easy to see that the total entropy is independent of the
choice $a$ or $b$, though its two components are not. As we will see below, we
are mostly interested in the configurational entropy, which from the above is
seen to depend on $a$. This is an unwanted behavior, since it makes $S(T)$
depend on the somewhat arbitrary parameter $a$. This unwanted property is
easily taken care of by normalizing the configurational entropy not by the
number of particles, but by the dimensionless volume $\overline{V}\equiv
V/a^{3}.$ It is easy to see that $\overline{S}(T)\equiv S(T)/\overline{V}$ is
independent of the arbitrary parameter $a$. In the following, we will always
interpret the discretized configurational entropy in this normalized sense.

It is now obvious that the entropy must be non-negative for equilibrium
states, since these states occur in Nature. This property need not hold for
metastable states, for which a negative entropy now will only indicate that
the corresponding state is \emph{not} realizable in Nature; it is no longer a
mere consequence of the continuum picture as above. An exact calculation, like
the one we carry out in this work, will evaluate the partition function
exactly \emph{without} using the saddle-point method. It will be seen that the
entropy is non-negative for the equilibrium state. We will also see that it
can become negative for the metastable state, where it will only imply that
such a state is not to be observed in Nature. The part of the metastable state
with non-negative entropy represents the metastable state that can occur in
Nature. Therefore, we will assume in the following that a discretization of
the two spaces has been carried out. In particular, a lattice model in which
the particle positions are restricted to be on the lattice sites is very
useful from this point of view. There is no kinetic energy in the lattice
model. Therefore, $S_{\text{KE}}(T)\equiv0.$ This leaves us with only the
configurational entropy, which represents the entire entropy in the lattice model.

\subsection{Other Common Definitions of the Configurational Entropy}

It should be clear from the above that the configurational entropy is a
concept that can only be introduced in classical statistical mechanics.
Unfortunately, there is no direct measurement of the configurational entropy
of a system.\ Thus, its experimental determination is a challenge. There is a
further complication in that there is a certain amount of confusion and/or
ambiguity about its definition. Our definition in (11) is the most natural
definition of the configurational entropy for which the requirement
$S(T)\geq0$ can be justified on the ground that the state must be realizable
in Nature. We will call this the \emph{thermodynamic principle of reality
}[8-10,12-13]. The configurational entropy represents that part of the total
entropy $S_{\text{T}}(T)$, which is due to the configurational degrees of
freedom only [3]. The remainder of the entropy $S_{\text{KE}}(T)$ is due to
the translational degrees of freedom and must be subtracted from the total
entropy to obtain $S(T)$. Thus, there is no ambiguity in its definition.

The ambiguity arises when we need to estimate $S(T)$ by subtracting the
contribution $S_{\text{KE}}(T)$ for the following reason. In a solid state
like the glass or CR, where the average particle positions are almost fixed,
$S_{\text{KE}}(T)$ is a significant part of the total entropy. The latter is
customarily known as the vibrational entropy in this context; we assume that
the container is stationary. The difference represents the configurational
entropy given by (11)$.$ If no discretization has been carried out, then
$S_{\text{KE}}(T)$ is given by (10) . This will surely create a problem at low
temperatures where $S_{\text{KE}}(T)$ and $S(T)$\ become negative. Thus,
discretization is necessary, and one needs to carry it out before
$S_{\text{KE}}(T)$ can be calculated. In that case, we need to normaize the
discretized$\ $configurational entropy $S(T)$ by the dimensionless volume
$\overline{V}$ as discussed above.

Traditionally, one treats the translational degrees of freedom approximately
in a different fashion, at least for CR. The discretization is enforced by
treating the resulting motion as giving rise to quantum vibrations about the
minimum of the CR configurational energy. [If we treat these vibrations
classically, it can be shown immediately that the kinetic energy contribution
to the entropy is precisely given by $S_{\text{KE}}(T).]$ Hence,
$S_{\text{KE}}(T)$ differs from the vibrational entropy by an amount exactly
equal to the configurational entropy, which need not be small. One further
assumes that the vibrational modes of SCL are similar to those in CR. One now
traditionally defines the SCL configurational entropy by [3]%

\[
S_{\text{con-}\mathrm{SCL}}(T)\simeq S_{\mathrm{ex,SCL}}(T)\equiv
S_{\text{T-}\mathrm{SCL}}(T)-S_{\text{T-}\mathrm{CR}}(T).
\]
Some workers identify the SCL configurational entropy by%

\[
S_{\text{con-}\mathrm{SCL}}^{\prime}(T)\simeq S_{\text{T-}\mathrm{SCL}%
}(T)-S_{\text{T-}\mathrm{GL}}(T),
\]
the excess SCL entropy over the extrapolated entropy of the corresponding
glass [43]. Workers using the potential landscape picture define the
configurational entropy to be the entropy $S_{\text{IS}}(E_{\text{IS}})$ of
inherent structures of a given potential energy $E_{\text{IS}}$ [37]$.$ To
obtain this entropy as a function of $T$, a connection between $E_{\text{IS}}$
and $T$ is required.

In this work, we will continue to use $S(T)$ to define the configurational
entropy. In a lattice model, which we consider in this work, we bypass the
complication due to $S_{\text{KE}}(T)$ all together since there is no kinetic
energy on a lattice. Hence, the entire entropy in the lattice model is purely
configurational. The role of the dimensionless volume $\overline{V} $ is
played by the number of lattice sites $N_{\text{L}}.$\ 

\subsection{Metastability Continuation}

A partitioning of the total PF into two or more PF's in CSM, such as in (5),
of decoupled degrees of freedom has an important consequence, which we discuss
below. We focus on (5), but the discussion is valid for any general
partitioning. The total entropy is a sum of the entropies from the two PF's.
The partitioning implies not only that the total entropy at a given
temperature is a sum of the entropy contribution due to $Z_{\text{KE}}$ and
that due to $Z$, but also that the two contributions are independent. In
particular, $S_{\text{KE}}(T)$ remains the same, whether we consider the
crystal or the supercooled phase, both of which can exist at the same
temperature below $T_{\text{M}}$. This is because CR and SCL are described by
$Z,$ and its appropriate modification to be described below, respectively.
Hence, we come to a very important conclusion:

In classical statistical mechanics, which is what is conventionally used to
analyze metastability, the contribution $S_{\text{KE}}(T)$ to the total
entropy $S_{\text{T}}(T)$ from the translational degrees of freedom is the
same for various possible phases like SCL or CR that can exist at a given
temperature, and is a function only of the temperature $T$.

Thus, from now on, we will only consider the configurational degrees of
freedom. Since the heat capacity is non-negative, $S(T),$ and $E(T)$ are
monotonic increasing function of $T$, and must reach their minimum values at
absolute zero. Hence, CR must be in the state with energy $E_{0},$ for which
we take $S_{\text{CR}}(T=0)=0.$ This requires taking $W(E_{0})=1.$ Thus,
$E_{0}$ \emph{sets the zero of the temperature scale in the system}.

At high temperatures ($T>T_{\text{M}}$), the system is an equilibrium liquid
(EL). At low temperatures ($T<T_{\text{M}}$), the system is an equilibrium
crystal CR. At $T_{\text{M}},$ the two equilibrium states are in equilibrium
where they have the same free energy. It is customary to distinguish the
disordered EL and the ordered CR by the use of the order parameter $\rho$
[44], which is conventionally defined in such a way that $\rho=0$ represents
the disordered phase, while $\rho\neq0$ represents the ordered phase. We will
now use the notion of the order parameter to describe the metastable state. We
will assume that the metastable state of interest is the one obtained under
infinitely slow cooling, but always ensuring that the CR is never allowed to
nucleate. We call such a metastable state a \emph{stationary metastable state}
(SMS). The most convenient way to describe SMS\ is by the use of a PF. For
this, we need to know the number of those configurations of energy $E$ that
yield SMS.\ 

We follow the well-established practice of describing these states by imposing
the constraint that the configurations must satisfy $\rho=0$ [44]. The number
of configurations $W(E)$ for a given energy $E$ can be partitioned into
$W_{\text{dis}}(E)$ and $W_{\text{ord}}(E),$ representing the number of
disordered ($\rho=0)$ and ordered ($\rho\neq0$) configurations, respectively.
Correspondingly, we have two entropy functions $S_{\text{dis}}(E)\equiv\ln
W_{\text{dis}}(E),$ $S_{\text{ord}}(E)\equiv\ln W_{\text{ord}}(E).$ At higher
energies, $S_{\text{dis}}(E)>S_{\text{ord}}(E)$. At lower energies,
$S_{\text{dis}}(E)<S_{\text{ord}}(E).$ From what has been said above, we have
$S_{\text{ord}}(E_{0})=0$ for the ideal crystal, which occurs at $T=0$. It is
usually the case that the glass has a higher energy ($E=E_{\text{K}}>E_{0}$)
at absolute zero. Thus, it appears likely that $S_{\text{dis}}(E_{\text{K}%
})=0.$ In this case, assuming that $E_{\text{K}}$\ is not a point of
singularity of $S_{\text{dis}}(E_{\text{K}}),$\ we perform a continuation of
$S_{\text{dis}}(E)$ to all energies $E\geq E_{0},$ which we also denote by
$S_{\text{dis}}(E)$, as there will be no confusion. We now use the extended
$S_{\text{dis}}(E),$ and $S_{\text{ord}}(E)$ to construct two configurational PF's%

\begin{equation}
Z_{\alpha}(T)=\sum_{E\geq E_{0}}W_{\alpha}(E)e^{-\beta E},\;\;\alpha
=\text{dis,ord.} \label{SMSPF}%
\end{equation}
\ For a macroscopic system, we have $Z(T)\cong Z_{\text{dis}}(T)$ for
$T>T_{\text{M}},$ and $Z(T)\cong Z_{\text{ord}}(T)$ for $T<T_{\text{M}}.$ We
now use the continuation of $Z_{\text{dis}}(T)$ below $T_{\text{M}}$ to
describe SMS, the continuation of EL. We can similarly use the continuation of
$Z_{\text{ord}}(T)$ above $T_{\text{M}}$ to describe the superheated CR above
the melting temperature. However, we are not interested in this continuation
here. We assume that $W_{\alpha}(E)\geq0$ in the following. As long
as\ $W_{\alpha}(E)\geq0$, $Z_{\alpha}(T)$ is a sum of positive terms. Hence,
both PF's will satisfy proper convexity properties. In particular, both will
yield non-negative heat capacity. Thus, we have a thermodynamically valid
description of CR\ and SMS at low temperatures.

As said above, $E_{0}$ sets the zero of the temperature scale. This is why we
perform the continuation $S_{\text{dis}}(E)$ to all energies $E\geq E_{0},$ so
that the same zero\ of the scale is common to both PF's in (15). This also
ensures that both states have the same common temperature $T_{\text{M}}$ at
the coexistence where $Z_{\text{dis}}(T_{\text{M}})$ $=Z_{\text{ord}%
}(T_{\text{M}}).$ The latter requirement is very important in that it shows
that SMS cannot be treated divorced from CR; both are required for enforcing a
proper and common temperature scale.

\subsection{Configurational Entropy Crisis}

For states to be observed in Nature, we require the number of microstates from
each PF in (15) not to be less than one: $W_{\alpha}(E)\geq1$. This also
ensures that $W_{\text{T}}\geq1,$ since we have already argued above\ that
$W_{\text{KE}}$ $\geq1$. This is another reason why there is no need to
consider translational degrees of freedom when discussing the entropy crisis.
We must only require that $W\geq1.$ For CR, which we know must exist in
Nature, we must certainly have $W\geq1.$ Recall that we have assumed $W(E_{0})=1.\ $

However, whether the entropies are non-negative is a separate requirement,
independent of the convexity properties. It is possible to have non-negative
heat capacity even if the entropy is negative. It is highly plausible from our
discussion above that\ $W_{\text{dis}}(E)<1$ for SMS for $E<E_{\text{K}}$.
Recall that it is likely that $W_{\text{dis}}(E)=1$ at $E_{\text{K}},$ and
must continue to decrease as $E$ decreases due to non-negative heat
capacity$.$\ 

Existence of a negative entropy causes no singularity or instability in the
corresponding PF. Thus, SMS PF can be continued all the way down to $T=0$.
Gujrati [12] has shown that the corresponding SMS free energy is identical
with that of CR. at $T=0,$ provided that $TS_{\text{SMS}}(T)\rightarrow0,$ as
$T\rightarrow0.$ However, at some positive temperature $T=T_{\text{K}},$ the
average $E_{\text{SMS}}(T)=E_{\text{K}};$ below\ $T_{\text{K}},$ SMS\ has a
negative entropy. This is not possible for any state that can exist in Nature.
Thus, this portion of SMS must be discarded and replaced by what is known as
the \emph{ideal glass} of energy $E=E_{\text{K}}$ and zero entropy. This is an
inactive state of zero heat capacity. Its necessity is not indicated by
anything that goes wrong with the PF, but by the additional requirement of
realizability condition that $S(T)\geq0.$ The ideal glass energy at absolute
zero remains $E_{\text{K}},$ which must represent a potential energy minimum,
just as the CR\ energy $E_{\text{0}}$ represents the global potential energy
minimum. Thus, at $T=T_{\text{K}},$ SMS is trapped in this minimum.

Under the assumption [12] that the glass has a higher energy ($E=E_{\text{K}}%
$) than the corresponding crystal ($E=E_{0}<E_{\text{K}}$)\ at absolute zero,
the entropy of the stationary SCL in any general system must necessarily
vanish at a positive $T_{\text{K}}.$ Gujrati provides two independent proofs
and substantiates the conclusions by two exact model calculations.

The following four observations by Gujrati [12] are relevant for our investigation.

\begin{itemize}
\item [G1.]The zero of the temperature scale is set by the potential energy
minimum $E_{0},$ and not by other minima like the potential energy minimum
$E_{\text{K}}.$

\item[G2.] The temperature scale of the disordered phase is fixed by its
coexistence with the ordered phase at the melting temperature, where both
phases have common free energy and temperature.

\item[G3.] The free energies of all phases, extended when necessary by
continuation without any regard to the entropy crisis, are equal at absolute
zero. Assuming $TS(T)\rightarrow0$ for all states as $T\rightarrow0$, this
results in the equal energy principle for all states at $T=0$.

\item[G4.] The ideal glass does not emerge directly in the statistical
mechanical description, but is put in by hand to avoid the entropy crisis.
\end{itemize}

\section{Model}

In order to study glass transition in molecular fluids, we need to consider a
solution of molecules and solvent particles. By a mere change of nomenclature,
we can use the same system to explain the behavior of a compressible system by
treating the solvent particles as representing voids. We are only going to
consider a lattice model here in which, as shown recently [45], excess or
exchange interaction between the solvent particles and the molecules is
sufficient to describe orientation-independent mutual interactions on a
lattice. This exchange interaction between molecules and solvent particles is
usually repulsive. Hence, at low temperatures, there would be phase separation
into a molecule-rich phase (a liquid) and molecule-poor phase (a gas). We are
only interested in the liquid phase, although our method allows us to capture
the solvent-rich phase also. The last is found to be always unstable. In a
lattice model, the ground state should be primarily controlled by the
interactions and not by the choice of the lattice; however, the latter may
have a secondary influence on the ground state. We wish to minimize the
effects of the lattice as much as possible to make our lattice results as
useful as possible for a real system. If the molecules are also taken as
monomeric particles that each occupy a site of the lattice, then such a system
is not a suitable candidate for studying glass transition for two reasons.
First, their ground state at absolute zero is completely determined by the
lattice chosen, since the interactions are no longer present as there are no
solvent molecules. The second reason is also very important. Simple liquids
with approximately spherical shape and pairwise interactions, such as
condensed rare gases or molten alkali halides are very hard to prepare into a
viscous liquid state. The simplest means to avoid this is to have anisotropic
particles with complex interactions so that the crystalline state is hard to
form. This will make it easier to supercool the equilibrium liquid EL below
the melting temperature. Dimers are the smallest molecules that (in the
absence of any solvent) can get into a unique ordered state not because of the
regular lattice structure, but because of physical interactions. Thus, we
consider dimers as the simplest molecules to minimize the effects of the
lattice on the ordered phase. Furthermore, dimers can also be thought of as
representing strongly correlated Cooper pairs in high-$T_{\text{c}}$
superconductors [46].

For the sake of convenience, therefore, we consider a model of classical
dimers with solvent on a square lattice, even though other lattices can be
considered. The presence of solvent, which can also be treated as void, is not
necessary for glass transition as shown recently [10]. Each solvent occupies a
site of the lattice. A dimer, on the other hand occupies two consecutive sites
and the intervening lattice bond between them. We restrict the excess
interaction (energy $\varepsilon)$ to a nearest-neighbor pair of a solvent and
an endpoint. It is easy to see that there is no one unique ground state at
$T=0.$ Thus, the mere use of the lattice does not produce an ordered
structure, which is very comforting for the reasons stated above. To create a
unique ground state to mimic a CR, we need to introduce additional
interactions between dimers. These are \emph{orientational interactions} and
our model is perfectly suited to investigate the importance of such
orientational interactions on the phase diagram. There is an orientational
interaction (energy $\varepsilon_{\text{p}})$ between a nearest-neighbor pair
of two unbonded dimer endpoints provided the corresponding dimers are
parallel. If this interaction is attractive, the ground state at $T=0$ is
columnar (Fig. 1a). If the interaction is repulsive, we introduce an
additional \textit{attractive} axial interaction $\varepsilon_{\text{a}}$
between the endpoints of two collinear dimers so that the ground state at
$T=0$ is staggered (Fig. 1b). Both ground states have a \emph{sublattice
structure }due to the squares falling into two distinct classes A and B, see
Fig. 1. Such ground states also play a central role in the short-ranged
resonant valence bond model of high-temperature superconductivity [46], where
the pair of parallel dimers (Fig. 1a) are said to resonate; the dimers in Fig.
1b are said to anti-resonate. The dimer interactions are induced by quantum
fluctuations. Anderson has hypothesized that the columnar phase with
resonating dimers is a good representation of pure La$_{2}$CuO$_{4}$ which is
an insulator.

\subsection{Partition Function Formalism}

The rigorous thermodynamic treatment of the model is carried out by using the
partition function formalism containing the solvent activity $\eta=\exp
(\mu/T),$ and the Boltzmann weights $w=\exp(-\varepsilon/T),w_{\text{p}}%
=\exp(-\varepsilon_{\text{p}}/T)$ and $w_{\text{a}}=\exp(-\varepsilon
_{\text{a}}/T).$ We set the Boltzmann constant $k_{\text{B}}=1$. We consider a
lattice containing finite and fixed $N_{\text{L}}$ lattice sites. The solvent
chemical potential $\mu$ is always kept negative ($\mu<0$) to insure a fully
dimer-packed ground state at absolute zero. The configurational PF is given as
follows:
\begin{equation}
Z=\sum\Omega(N_{0},N_{\text{p,}}N_{\text{a}},N_{\text{c}})\eta^{N_{0}%
}w_{\text{p}}^{N_{\text{p}}}w_{\text{a}}^{N\text{a}}w^{N_{\text{c}}%
},\label{44}%
\end{equation}
where $N_{0}$ is the number of the solvent molecules, $N_{\text{p}}$ the
number of the nearest-neighbor endpoint contacts between dimers that resonate,
$N_{\text{a}}$ the number of nearest-neighbor axial contacts between the
endpoints of collinear dimers (pointing in the same direction), $N_{\text{c}}$
the number of nearest-neighbor solvent-endpoint contacts, and $\Omega$ the
number of distinct configurations for a given $N_{0},$ $N_{\text{p}}$,
$N_{\text{a}},$ and $N_{\text{c}}$. The sum is over $N_{0},$ $N_{\text{p}},$
$N_{\text{a}}$ and $N_{\text{c}}$, consistent with a fixed and finite
$N_{\text{L}}$. The densities in the model, to be denoted by $\phi_{k}$, are
defined by the limiting ratios $N_{k}/N_{\text{L}}$ \emph{per site} as
$N_{\text{L}}\rightarrow\infty;$ here $k=0,$p$,$a, and c. The ground state at
$T=0$ is determined by whether $\varepsilon_{\text{p}}<0$ (attractive) or
$\varepsilon_{\text{p}}>0$ (repulsive).\ For the attractive case, we set
$\varepsilon_{\text{a}}=0$ for simplicity, as its presence does not change the
topology of the phase diagram. As shown elsewhere [47], the adimensional free energy%

\[
\omega=\lim_{N_{\text{L}}\rightarrow\infty}(1/N_{\text{L}})\ln Z
\]
represents the osmotic pressure across a membrane permeable to the dimers, but
not solvent. The reduced pressure $\pi_{0}\equiv Pv_{0},$ where $v_{0}$ is the
volume of a lattice site, is given by
\[
\pi_{0}=T\omega-\mu.
\]
\ The equilibrium state must have the highest pressure among all possible
states obtained at given $T$, $w_{\text{p}}$, $w_{\text{a}}$ and $\mu$
[47].\ The total energy per site $E$\ is given by
\begin{equation}
\beta E=-(\phi_{\text{p}}\ln w_{\text{p}}+\phi_{\text{a}}\ln w_{\text{a}}%
+\phi_{\text{c}}\ln w).\label{energy}%
\end{equation}
The entropy per site $S$ is calculated through the Legendre transform of the
adimensional osmotic pressure:
\begin{equation}
S=\omega+\beta E-\phi_{0}\ln\eta,\label{entropy}%
\end{equation}
where $\phi_{0}$ is the solvent density. Since we are considering a
configurational PF, $S$ is purely configurational. (From now on, $S$ and $E$
will represent the configurationl entropy and energy per site and not the
entropy and energy of the entire system.) At $T=0$, $\phi_{\text{c}}=0.$ The
maximal values of $\phi_{\text{p}},\phi_{\text{a}}$ are easily seen to be 1,
and 1/2 (see insets in the Figs. 3c and 7c), respectively. The density
$\phi_{\text{e}}$ of nearest-neighbor endpoint pairs due to dimers neither
parallel nor collinear has its maximum density 3/2 in the configuration in
which dimers form staggered steps. We consider the two cases $\varepsilon
_{\text{p}}<0$ (attractive) and $\varepsilon_{\text{p}}>0$ (repulsive)
separately.\ The number of endpoint contacts between resonating dimers is
twice the number of resonating dimer pairs.\ Thus, if $\phi_{\text{RD}}$
denotes the density of pairs of resonating dimers, then $\phi_{\text{RD}}%
=\phi_{\text{p}}/2.$

\section{Husimi Tree Approximation}

In order to solve the model \emph{exactly}, we replace the square lattice by a
site-sharing \ Husimi cactus shown in Fig. 2, obtained by connecting two
squares at each site. This is the only approximation we make. The cactus can
be thought of as a checker-board version of the square lattice, representing
squares of a given color [9], so that the squares of the other color are
missing. However, a pair of dimers on the cactus that would have belonged to a
missing square on the original square lattice is counted as a parallel pair on
the cactus. Husimi tree incorporates local loop structure of the regular
lattice. It is more advantageous than the Bethe lattice as far as the effect
of local correlations is concerned, since the Bethe lattice approximation
disregards presence of any topological loops. The model is solved exactly on
the cactus by \emph{recursive technique}, which results in RR's, as shown
elsewhere [48]. Because of its exactness, the calculation respects

\begin{itemize}
\item [(i)]all local (such as gauge) and global symmetries, in contrast to the
conventional mean-field solution which is known to violate local symmetries, and

\item[(ii)] thermodynamics is never violated.
\end{itemize}

This makes the present approach superior to other methods of constructing
mean-field solutions based on random mixing approximation [48], because more
local correlations are taken into account. The advantage of using the
recursive lattices for the thermodynamic calculations was pointed by Gujrati
[48] in connection with a simple trick that allows the calculation of the free
energy $\omega$, and various densities exactly.

\subsection{Recursion Relations}

The lattice sites are indexed by a level index $m$, which increases as we move
away from the center of the cactus [8]. If the base site, which is close to
the cactus center, of a given square is labeled $m$, then the other three
sites of the square (the two intermediate sites connected to the base and the
peak site connected to the intermediate sites) are indexed ($m+1$). This
square is indexed as $m$th square and is connected to three ($m+1$)th squares.
Another index $\alpha$, called the directional index, is introduced to capture
the staggered phase on the Husimi tree. We subdivide the lattice sites in four
different classes representing the directions on the lattice: up, down, left,
and right; see Fig. 1. The squares that lay above their base site are labeled
$\mathcal{U}$ (up)$,$ below -$\,\mathcal{D}$ (down)$,$ to the left -
$\mathcal{L}$, and to the right - $\mathcal{R}$ . A square\ of type
$\mathcal{U}$ with its base site at the $m$th level is connected to
$\mathcal{L}$, $\mathcal{U}$, and $\mathcal{R}$ (clockwise with respect to the
base site) squares with their base sites at the ($m+1$)th level. Similarly, a
$\mathcal{D}$ square is connected to the $\mathcal{R}$, $\mathcal{D}$, and
$\mathcal{L}$ squares, a $\mathcal{L}$ square is connected to the
$\mathcal{D}$, $\mathcal{L}$, and $\mathcal{U}$ squares, and a $\mathcal{R}$
square is connected to the $\mathcal{U}$, $\mathcal{R}$, and $\mathcal{D}$
squares on the upper level. The directional index $\alpha$ of the square will
also be associated with its base site. As a convention, the directional index
of the peak site\ of a given square is the same as that of its base site.
There are two equivalent ways to view the entire cactus with respect to its
origin at $m=0$ [49]. In one view, the origin is taken to be connected to four
squares labeled $\mathcal{L}$, $\mathcal{U}$, $\mathcal{R}$, and $\mathcal{D}%
$, as shown in the Fig. 2. In the other view, the origin is taken to be
connected to only two squares labeled $\mathcal{L}$, and $\mathcal{R}$, or
$\mathcal{U}$ and $\mathcal{D}$. The former is a symmetric description of the
cactus, but more cumbersome, while the latter is an asymmetric description,
but easier for calculations. As shown elsewhere [49], both are equivalent in
all their consequences.

There are five possible states for a site on the lattice. It is either
occupied by a solvent (to be denoted by 0) or by a dimer endpoint with the
dimer pointing along one of the four bond directions that we denote by
horizontal going up, i.e. pointing away from the origin (h$_{\text{u}}$), or
down, i.e. pointing towards the origin (h$_{\text{d}}$), and vertical going up
or down (v$_{\text{u}},$v$_{\text{d}}$); see Fig. 1 for the horizontal (h) and
vertical (v) directions. The dimers align themselves on the square lattice in
the completely ordered states along a single direction, vertical or
horizontal. We introduce the partial partition functions (PPF) $Z_{m}%
^{(\alpha)}(i)$, given that the site on the level $m$ is in the state
$i=0,$h$_{\text{u}},$h$_{\text{d}}$,v$_{\text{u}},$v$_{\text{d}},$ and has the
directional index $\alpha=$ $\mathcal{U}$,$\mathcal{D}$,$\mathcal{L}%
$,$\mathcal{R}.$ It represents the contribution to the total partition
function from the part of the tree above $m$-th level site in the ($i,\alpha
$)$-$state. We then obtain the RR's between $Z_{m}^{(\alpha)}(i)$ at level
$m\ $and the PPF's at the higher level $(m+1)$ by considering all
possibilities and the local statistical weights of the $(m+1)$th square. In
the most general case, the RR's for the PPF's are given by a cubic relation
for a square Husimi cactus:
\begin{equation}
Z_{m}^{(\alpha)}(i)=\sum w_{jkp}^{(\gamma\delta\alpha)}Z_{m+1}^{(\gamma
)}(j)Z_{m+1}^{(\alpha)}(k)Z_{m+1}^{(\delta)}(l), \label{RR}%
\end{equation}
where ($j,\gamma),$ ($k,\alpha),$ and ($l,\delta$) are the states of the three
sites at the level ($m+1)$ (in the clockwise direction from the base site) for
each allowed configuration inside the $m$th level polygon and the base site is
in the state ($i,\alpha$). The local statistical weight is given by
$w_{jkl}^{(\gamma\alpha\delta)}$. The directional indices are $(\gamma
\alpha\delta)=(\mathcal{LUR)},(\mathcal{RDL}),(\mathcal{DLU}),$ and
$(\mathcal{URD})$ for $\alpha=\mathcal{U},\mathcal{D},\mathcal{L},$and
$\mathcal{R},$ respectively. We then use%

\begin{equation}
B_{m}^{(\alpha)}\equiv\sum_{i}Z_{m}^{(\alpha)}(i), \label{Amplitude}%
\end{equation}
in which the sum is over all states $i$, to introduce the ratios%

\begin{equation}
x_{i,m}^{(\alpha)}\equiv Z_{m+1}^{(\alpha)}(i)/B_{m+1}^{(\alpha)}.
\label{Ratios}%
\end{equation}
It is obvious that the ratios always satisfy the sum rule%

\begin{equation}
\sum_{i}x_{i,m}^{(\alpha)}=1. \label{sumrule}%
\end{equation}
The above RR's among the partial PF's is then converted into RR's relate
$x_{i,m}^{(\alpha)}$ with the ratios at the higher level $(m+1)$:
\begin{equation}
x_{i,m}^{(\alpha)}=P_{i,m+1}^{(\alpha)}/Q_{m+1}^{(\alpha)}, \label{xRR}%
\end{equation}
where $P_{i,m}^{(\alpha)}$ are various polynomials given by the right-hand
side of (19) in which the PPF's are replaced by the corresponding ratio given
in (21) and
\begin{equation}
Q_{m}^{(\alpha)}=\underset{i}{\sum}P_{i,m}^{(\alpha)}. \label{Denominator}%
\end{equation}
The explicit expressions for $P_{i,m}^{(\alpha)}$ are given in the Appendix.

\subsection{Fix-Point Solutions}

The fix-point (FP) solution of the RR's describe the behavior in the interior
of an infinite Husimi cactus. The FP solution can be obtained numerically or
analytically (when possible). There are two different kinds of FP solutions.
In the 1-cycle FP solution, the ratios remain the same at consecutive levels.
We denote the FP values of the ratios by $\{x_{i}^{(\alpha)}\}\equiv
\{s^{(\alpha)},h_{\text{u}}^{(\alpha)},h_{\text{d}}^{(\alpha)},v_{\text{u}%
}^{(\alpha)},v_{\text{d}}^{(\alpha)}\}$. We wish to emphasize that the
italicized quantities $h_{\text{u}}^{(\alpha)},$ etc. represent the values of
the FP, and should not be confused with the states (h$_{\text{u}}$,$\alpha$),
especially when the description does not require the use of the directional
index $\alpha$, which is very much possible as we will see below. The
italicized quantities will always represent FP values. In the 2-cycle FP
solution, the ratios alternate between two successive levels, which are
referred to as \emph{even} and \emph{odd} in the following. The ratios on even
levels are denoted without prime, and on odd levels with prime. The free
energy is calculated using the method originally proposed by Gujrati [48] for
the 1-cycle FP and its extension given in [9] for the 2-cycle FP. The FP
solution that maximizes the osmotic pressure over a temperature range
represents the equilibrium state over that range. This exact solution is taken
as the \emph{approximate} theory for the square lattice. The approach allows
us to describe both the ordered, i.e. the crystal (CR) phase at low
temperatures and the disordered, i.e. the equilibrium liquid (EL) phase at
high temperatures. In addition, an intermediate phase (IP) with intermediate
orientational order [more ordered with respect to EL and less ordered with
respect to CR] is discovered by the analysis and is involved in a
liquid-liquid (L-L) transition induced by orientational interactions; see
below for a complete description. True equilibrium states are those that
minimize the relevant free energy. Abandoning this minimization principle
allows us to obtain metastable states by continuing various solutions [12].
The continuation of EL to lower temperatures describes the supercooled liquid
(SCL), which as we show here exhibits the entropy crisis at $T=T_{\text{K}}.$
The disordered liquid EL for the attractive case and SCL for the repulsive
case undergo a transition to IP. The continuation of IP to lower temperatures
also exhibits an entropy crisis of its own at a temperature $T=T_{\text{K}%
}^{\prime}$ that is usually different from $T_{\text{K}}$. We study various
contact densities in CR, IP and EL, and their continuation. The densities
contributing to the energy of different states, upon continuation, approach
their corresponding values for CR as $T\rightarrow0$. Thus, all phases will
have \emph{identical} free energies at $T=0$ if they can exist there. This is
consistent with the claim G3 [12]. This equality [8,12] ensures that each of
the two metastable states obtained by continuing EL and IP will exhibit
entropy crisis.

\section{Results}

We consider the cases $\eta=0$, and $\eta>0$, separately as we are able to
obtain analytical solutions for the former case. For the case $\eta=0$,
$\omega$ is related to the Helmholtz free energy $F(T)$: $F(T)=-T\omega$ [8].
For $\eta>0,$ $\pi_{0}$ is related to the thermodynamic potential
$F(T)=-\pi_{0}.$ To cover both cases together we define the following shifted
thermodynamic potential $\overline{F}$:
\[
\overline{F}(T)\equiv F(T)-F(0)=-T\omega-E_{0},
\]
where $E_{0}$ is the energy of the perfect CR at $T=0$. It has the property
that it vanishes at $T=0$. The vanishing of entropy corresponds to the maximum
in $\overline{F}(T).$

\subsection{Disordered Phase:\ EL\ and SCL}

The disordered phase is the continuation of the phase at infinite
temperatures, where the correlations are minimal. This phase is described by a
1-cycle FP solution.

\subsubsection{Absence of Solvent}

We study both attractive and repulsive cases together. The relevant FP
solution corresponds to having complete equivalence among all quantities with
different directional indices. Thus, we will suppress the directional index
$\alpha$ ($x_{i}^{(\mathcal{R})}$ = $x_{i}^{(\mathcal{L})}$ = $x_{i}%
^{(\mathcal{D})}$ = $x_{i}^{(\mathcal{U})}\equiv x_{i}$) at present. The RR's
for such a FP solution are the following:%

\begin{align}
Qh_{\text{d}}\text{{}}  &  =\text{(}w_{\text{a}}w_{\text{p}}\text{)}%
^{2}h_{\text{u}}^{3}+3w_{\text{a}}w_{\text{p}}h_{\text{u}}^{2}v_{\text{u}%
}+(1+2w_{\text{a}}w_{\text{p}})v_{\text{u}}^{2}h_{\text{u}}\label{1}\\
&  +w_{\text{a}}w_{\text{p}}v_{\text{u}}^{3}+w_{\text{a}}h_{\text{d}}%
^{2}h_{\text{u}}+h_{\text{d}}^{2}v_{\text{u}}+w_{\text{p}}h_{\text{u}%
}v_{\text{d}}^{2}+v_{\text{u}}v_{\text{d}}^{2},\nonumber\\
Qv_{\text{d}}  &  =w_{\text{a}}w_{\text{p}}h_{\text{u}}^{3}+3w_{\text{a}%
}w_{\text{p}}v_{\text{u}}^{2}h_{\text{u}}+(1+2w_{\text{a}}w_{\text{p}%
})h_{\text{u}}^{2}v_{\text{u}}\nonumber\\
&  +(w_{\text{a}}w_{\text{p}})^{2}v_{\text{u}}^{3}+h_{\text{d}}^{2}%
h_{\text{u}}+w_{\text{p}}h_{\text{d}}^{2}v_{\text{u}}+h_{\text{u}}v_{\text{d}%
}^{2}+w_{\text{a}}v_{\text{u}}v_{\text{d}}^{2},\nonumber\\
Qv_{\text{u}}  &  =v_{\text{d}}(w_{\text{p}}h_{\text{u}}^{2}+2v_{\text{u}%
}h_{\text{u}}+w_{\text{a}}v_{\text{u}}^{2}+w_{\text{p}}^{2}v_{\text{d}}%
^{2}),\nonumber\\
Qh_{\text{u}}  &  =h_{\text{d}}(w_{\text{p}}v_{\text{u}}^{2}+2v_{\text{u}%
}h_{\text{u}}+w_{\text{a}}h_{\text{u}}^{2}+w_{\text{p}}^{2}h_{\text{d}}%
^{2}).\nonumber
\end{align}
The polynomial $Q,$ see (24), is given by the sum of the right-hand sides in
the above equation because of the sum rule (22) for the ratios,\ where we must
recognize that $s$ is zero because of $\eta=0.$

The osmotic pressure for this solution is calculated from general expressions,
which follows from the Gujrati trick [48]. The central square is conneted to 4
branches extending in different directions. Recalling that the index $\alpha$
is suppressed in the quantity $B_{m}^{(\alpha)},$ we find that the total
PF\ of the entire lattice is given by
\[
Z_{0}=B_{0}^{4}QQ_{0},
\]
where
\[
Q_{0}=2v_{\text{u}}v_{\text{d}}+2h_{\text{u}}h_{\text{d}}+s^{2}/\eta.
\]
We now imagine taking out the 5 squares at the origin, which leaves behind 9
smaller branches of the cactus, from which we construct 3 smaller lattices by
connecting 4 branches together each. The total PF of each of the smaller
lattices is given by
\[
Z_{1}=B_{1}^{4}QQ_{0},
\]
so that [48]
\[
\omega=(1/4)\ln(Z_{0}/Z_{1}^{3}).
\]
Since $B_{0}=B_{1}^{3}Q,$ see (20) and (21), we finally have
\begin{equation}
\omega=(1/2)\ln\left(  Q/Q_{0}\right)  .\nonumber
\end{equation}

We can also calculate the densities of various types of contacts. For this, we
consider the single square at the origin and consider all of their
possibilities that contribute to the quantities of interest. Dividing this
contribution with the total PF $Z_{0}$ yields the density per site as
$\phi_{\lambda}=\Phi_{\lambda}/QQ_{0},$ $\lambda=$p,a, where%

\begin{align}
\Phi_{\text{p}}  &  =w_{\text{p}}(2w_{\text{a}}h_{\text{u}}^{2}v_{\text{u}%
}^{2}+v_{\text{d}}^{2}h_{\text{u}}^{2}+h_{\text{d}}^{2}v_{\text{u}}%
^{2}+2w_{\text{a}}v_{\text{u}}^{3}h_{\text{u}}+2w_{\text{a}}v_{\text{u}%
}h_{\text{u}}^{3})\label{5}\\
&  +w_{\text{p}}^{2}(w_{\text{a}}^{2}v_{\text{u}}^{4}+w_{\text{a}}%
^{2}h_{\text{u}}^{4}+h_{\text{d}}^{4}+v_{\text{d}}^{4}),\nonumber
\end{align}%
\begin{align}
\Phi_{\text{a}}  &  =w_{\text{a}}(2w_{\text{p}}v_{\text{u}}h_{\text{u}}%
^{3}+v_{\text{d}}^{2}v_{\text{u}}^{2}+2w_{\text{p}}v_{\text{u}}^{3}%
h_{\text{u}}+2w_{\text{p}}v_{\text{u}}^{2}h_{\text{u}}^{2}+h_{\text{d}}%
^{2}h_{\text{u}}^{2})+\label{6}\\
&  +(w_{\text{a}}w_{\text{p}})^{2}(h_{\text{u}}^{4}+v_{\text{u}}%
^{4}).\nonumber
\end{align}

There is an additional symmetry in the FP solution of interest for the
disordered phase. This symmetry corresponds to having
\begin{align}
h_{\text{u}}  &  =v_{\text{u}}\equiv a,\label{2}\\
h_{\text{d}}  &  =v_{\text{d}}\equiv b.\nonumber
\end{align}
The resulting system of FP equations are
\begin{align*}
b  &  =\left(  a^{3}K_{1}+ab^{2}K_{2}\right)  /Q,\\
a  &  =\left(  ba^{2}K_{2}+b^{3}w_{\text{p}}^{2}\right)  /Q,\\
Q  &  =2\left(  a^{3}K_{1}+ab^{2}K_{2}+ba^{2}K_{2}+b^{3}w_{\text{p}}%
^{2}\right)  ,
\end{align*}
where
\begin{align*}
K_{1}  &  =(w_{\text{a}}w_{\text{p}})^{2}+6(w_{\text{a}}w_{\text{p}})+1,\\
K_{2}  &  =w_{\text{a}}+w_{\text{p}}+2,
\end{align*}
can be solved analytically making the substitution $r=b/a$:
\begin{equation}
r=(K_{1}/w_{\text{p}}^{2})^{1/4}. \label{3}%
\end{equation}
Using the sum rule $2a+2b=1,$ which follows from (22), we eventually have
\begin{align*}
Q^{\text{dis}}  &  =ab(K_{2}+\sqrt{K_{1}w_{\text{p}}^{2}}),\\
Q_{0}^{\text{dis}}  &  =4ab.
\end{align*}

We finally find for the disordered phase
\begin{align}
\omega^{\text{dis}}  &  =\frac{1}{2}\ln\left(  K_{2}+\sqrt{K_{1}w_{\text{p}%
}^{2}}\right)  -\ln2,\label{7}\\
\phi_{\text{p}}^{\text{dis}}  &  =\frac{3w_{\text{p}}w_{\text{a}}+w_{\text{p}%
}r^{2}+2(w_{\text{p}}w_{\text{a}})^{2}+2w_{\text{p}}^{2}r^{4}}{2r^{2}%
(K_{2}+\sqrt{K_{1}w_{\text{p}}^{2}})},\nonumber\\
\phi_{\text{a}}^{\text{dis}}  &  =\frac{3w_{\text{a}}w_{\text{p}}+w_{\text{a}%
}r^{2}+(w_{\text{a}}w_{\text{p}})^{2}}{2r^{2}(K_{2}+\sqrt{K_{1}w_{\text{p}%
}^{2}})},\nonumber\\
S_{\text{dis}}  &  =\omega^{\text{dis}}-\phi_{\text{p}}^{\text{dis}}\ln
w_{\text{p}}-\phi_{\text{a}}^{\text{dis}}\ln w_{\text{a}}.\nonumber
\end{align}
The solution described above represents the equilibrium state at high
temperatures. Thus, we identify it as the disordered liquid phase stable at
high temperatures, see below for a complete description.

At this point we refer to the famous calculations by Kasteleyn [50] and by
Fisher [51] who provided the exact value for the number of fully-packed dimer
packings (no solvent) on a square lattice, $\varphi_{\text{exact}}=1.7916$
where $\varphi$ is the molecular freedom per site, $\varphi=\exp(2S)$. The
high-temperature limit of $S_{\text{dis}}$ in (30) represents the number of
configurations for weakly interacting dimers on Husimi cactus. Taking the
limit $w_{\text{a}}\rightarrow1,$ $w_{\text{p}}\rightarrow1$ we find
$\varphi_{\text{Husimi}}=1.7071.$ To appreciate the improvement due to the use
of the Husimi cactus over the Bethe lattice we compare this with
$\varphi_{\text{Bethe}}=1.6875$ obtained by Chang [52].

\subsubsection{Presence of Solvent}

We investigate the effect of solvent by letting $\eta\neq0.$ We now consider
nearest-neighbor solvent -dimer end-point interactions, which we take not do
depend on the dimer orientation. These are ordinary isotropic interactions
with energy given by the standard exchange or excess expression:
$\varepsilon=e_{0\text{m}}-1/2(e_{00}+e_{\text{mm}})=-1/2e_{\text{mm}}$, where
$e_{0\text{m}},$ $e_{00}$, and $e_{\text{mm}}$\ are direct solvent-monomer,
solvent-solvent, and monomer-monomer interaction energies [45]. Apart from
$\varepsilon$, we have the chemical potential $\mu$ to control the amount of
solvent particles in the system. We are no longer able to find the
FP\ solution analytically, and are forced to use numerical methods to find it.
The presence of solvent modifies the RR's in (25), since we need to also
incorporate the state $i=0$ describing the solvent presence at each level. The
general RR's in the Appendix can be used to construct the 1-cycle RR's. The
results of this numerical solution is presented below when we discuss the
entire phase diagram.

\subsection{Ordered Phases CR and IP in the Attractive Case}

In case when parallel-dimer configurations become more favorable as
$T\rightarrow0,$ the ground state is the columnar phase.\ There is no need for
the axial interactions to attain the complete order at $T=0$; therefore first
we set $\varepsilon_{\text{a}}=0$, and consider the effect of $\varepsilon
_{\text{a}}<0$ later. Attractive axial interaction only enhances the chance of
forming the ordered state. The symmetry of the ground state does not require
using directional indices thereby simplifying our RR's. In addition to the
disordered equilibrium liquid EL discussed above and stable at high
temperatures, our solution captures two ordered phases: columnar state CR,
which is the equilibrium state at low temperatures, and a phase IP with an
intermediate order, which is the equilibrium state at intermediate
temperatures before EL becomes the equilibrium state.

\subsubsection{Absence of Solvent}

At first, we study the pure dimer system with $\eta=0$. Three phases are
distinguished by the symmetries of their FP structures. The EL solution was
already described in previous section, where it was found that $h\equiv
h_{\text{u}}+h_{\text{d}}$ $=$ $v\equiv v_{\text{u}}+v_{\text{d}}=a+b=1/2$ -
the numbers of vertical and horizontal dimers are the same. The perfect
columnar CR ($T=0$) on the cactus has each A square surrounded by B squares,
and vice versa - consider the checker-board version of the A-B sublattice
structure for squares in Fig. 1a. This property is captured by applying a
2-cycle scheme when the FP conditions for even and odd lattice sites are
attained separately. We take the convention that even (odd) squares on the
Husimi tree have even (odd) base sites. In the case when the dimers in the
perfect CR occupy even squares and are oriented in the horizontal direction
(horizontally resonating dimers), we have $h_{\text{u}}=1$, $v_{\text{u}}=0,$
$h_{\text{d}}^{\prime}=1,$ $v_{\text{d}}^{\prime}=0$. The perfect CR with
dimers resonating vertically $(v_{\text{u}}=1$, $h_{\text{u}}=0,$
$v_{\text{d}}^{\prime}=1,$ $h_{\text{d}}^{\prime}=0)$ or resonating dimers
occupying odd polygons are disjoint from the above horizontally resonating CR
phase and does not have to be considered separately. As the temperature is
raised, CR begins to distort such that $h_{\text{u}},$ and $h_{\text{d}%
}^{\prime}$ decrease and $v_{\text{u}},$ and $v_{\text{d}}^{\prime}$ increase.
On the other hand, IP is given by a solution which satisfies $h_{\text{u}}=$
$v_{\text{u}}=$ $h_{\text{d}}^{\prime}=$ $v_{\text{d}}^{\prime}=1/2$ and which
is \emph{independent} of temperature. The 2-cycle structure of the solution
suggests that dimers continue to resonate locally occupying even squares, but
resonating pairs on average have no preference whether to orient horizontally
or vertically. Therefore, we can say that the IP is characterized by the
positional order but no orientational order, which is the general definition
of a \emph{plastic crystal}.

The perfect CR that has $T\omega^{\text{ord}}=-\varepsilon_{\text{p}}$,
$S_{\text{ord}}=0$, $\phi_{\text{p}}^{\text{ord}}=1$, $\phi_{\text{a}%
}^{\text{ord}}=1/2$ is thermodynamically stable at temperatures close to zero
as expected. In the following we assume that all dimers are horizontal. As the
CR heats up, resonating dimer pairs begin to resonate in the vertical
direction maintaining their positions within even squares with the result that
$S_{\text{ord}}$ increases, and $\phi_{\text{p}}^{\text{ord}}$ and
$\phi_{\text{a}}^{\text{ord}}$ decrease. At some point, CR undergoes a
continuous transition and a new state IP appears; the transition point is
shown in Fig. 3 at $T_{\text{c}}\cong1.7$. In IP, the densities of resonating
dimers in the two directions remain \emph{equal} to each other. The density of
endpoint contacts due to dimers resonating within A (or B) squares only
$\phi_{\text{p,res}}^{\text{IP}}=1/2$ remains independent of the temperature,
while $\phi_{\text{p}}^{\text{IP}}$ continues to decrease slowly. We also have
($K_{1}=1+6w_{p}+w_{p}^{2})$%
\begin{gather*}
\phi_{\text{a}}^{\text{IP}}=(3w_{\text{p}}+w_{\text{p}}^{2})/2K_{1},\text{
\ \ }\phi_{\text{p}}^{\text{IP}}=1/2+\phi_{\text{a}}^{\text{IP}},\\
\omega^{\text{IP}}=(1/4)\ln\left(  K_{1}w_{\text{p}}^{2}/4\right)  .
\end{gather*}

Figure 3 shows the equilibrium states by solid lines, EL-metastable extension
SCL at lower temperatures by the dashed line, IP-metastable extensions at
higher and lower temperatures by dash-dotted lines. The equilibrium status is
resolved by comparing the free energy $F(T)=-T\omega$ for all solutions
obtained at a given temperature as shown in Fig. 3a. We also present the
specific heat curves, $C$, in the same figure but using the right axis. Figure
3c shows the configurational entropy, and the contact densities $\phi
_{\text{a}}$ and $\phi_{\text{p}}$ are given in the inset. Note that the
configurational entropy becomes zero at the point where $F(T)$ is maximum.
Below the maximum, the entropy becomes negative and, therefore, this portion
of the extension cannot represent any realizable state. The extensions are
shown by dotted lines. Thus, the maximum in $F(T)$ locates the Kauzmann
temperature. This is our main result. We also note that both SCL and the
supercooled IP states exhibit the entropy crisis. The solutions are obtained
below correspond to negative entropies According to ideas traced back to times
of entropy crisis discovery [1,16], a metastable curve must be
\emph{terminated} at its Kauzmann point where the system undergoes an
\emph{ideal glass} transition, the ideal glass being a frozen and unique
disordered structure. Note that the transition is not dictated by the
statistical mechanical treatment, but by an additional requirement that the
configurational entropy be non-negative; see G4 above. The emergence of an
ideal glass cannot be verified experimentally due to tremendous slowing down
below the experimentally observed glass transition. We emphasize here that the
above entropy crisis has been demonstrated for the first time in an explicit
calculation for a model of molecular liquid.

The EL undergoes a first-order transition to IP at the melting temperature,
$T_{\text{M}}\cong3.6$ in Fig. 3. The continuation of EL below $T_{\text{M}}$
gives rise to SCL, whose entropy vanishes near $T_{\text{K}}\cong0.64,$ thus
establishing the existence of an entropy crisis in small molecules. Similarly,
the continuation of IP below $T_{\text{c}}$ describes its metastable state
whose entropy also vanishes near $T_{\text{K}}^{^{\prime}}\cong0.44$, thus
giving rise to another entropy crisis. Our exact calculation also shows that
the contact densities of both metastable states approach the values for the
ideal crystal as $T\rightarrow0$, namely $\phi_{\text{a}}^{\kappa}%
\rightarrow1/2$, $\phi_{\text{p}}^{\kappa}\rightarrow1$, where $\kappa=$ dis,
IP, and CR. Equality of $\phi_{\text{p}}^{\kappa}$ is consistent with the
energy equality at $T=0;$ see G3 above. The equality of $\phi_{\text{a}%
}^{\kappa}$ is due to the fact that it is determined uniquely by
$\phi_{\text{p}}^{\kappa}$ at $T=0$. The entropies of the metastable branches
approach negative values at $T=0$: $S_{\text{IP}}\rightarrow-\ln\sqrt{2}$, and
$S_{^{\text{dis}}}\rightarrow-\ln2$.

The differences between the three phases in the attractive case can be
summarized as below. Let $h\equiv h_{\text{u}}+h_{\text{d}},=$ and $v\equiv
v_{\text{u}}+v_{\text{d}}$ at each level, whether even or odd. (For odd
levels, we have primed quantities$.)$ The calculation shows that

\begin{itemize}
\item [(1)]EL has the 1-cycle structure and has $h=v$ at each level.

\item[(2)] IP has the 2-cycle structure but still has $h=v$ at each level.

\item[(3)] CR has the 2-cycle structure but does not satisfy $h=v$ at each level.
\end{itemize}

\subsubsection{Presence of Solvent}

The presence of solvent eventually makes melting transition to become
second-order. Figure 4 includes the curves for a typical solvent density
dependence shown by the upper curves, and the ratio of the density of
solvent-solvent contacts to the solvent density,
\[
D_{0}\equiv\phi_{00}/\phi_{0},
\]
shwon by the lower curves. The inset shows the entropy curve for this case.
The ratio $D_{0}\equiv N_{00}/N_{0}$ representing the number of
solvent-solvent contacts per solvent is a measure of solvent localization in
analogy with holon localization in the quantum dimer model. If the solvent
particles were randomly distributed, then $\phi_{00}=2\phi_{0}^{2},$ so that
$D_{0}=2\phi_{0}.$ If the solvent particles were paired as two
nearest-neighbor solvent particles (dimerized solvent) so as to replace a
dimer, then $\phi_{00}=\phi_{0}/2$, and $D_{0}=1/2.$ If the solvent particles
appear as a cluster of four to replace two dimers inside a square, then
$D_{0}=1,$ and so on. Thus, $\phi_{00}/\phi_{0}$ provides us with the
information about the nature of their distribution. We expect the random
distribution at very high temperratures. From Fig. 4, we see that $D_{0}$ is
sligfhtly larger than $\phi_{0}$ at higher temperaturers, implying that the
solvent distribution is still highly correlated. At low temperatures, we find
that $D_{0}^{\text{CR}}=D_{0}^{\text{IP}}=1/2$, indicating that solvents
appear in the form of a dimerized solvent to replace a single dimer. In
contrast, $D_{0}^{\text{SCL}}$ approaches zero, implying that the solvents are
not dimerized in it. This is an interesting\ observation in view of the fact
that all have vanishingly small solvent density near absolute zero. At higher
temperatures, $\phi_{0}^{\text{SCL}}$ is much higher than $\phi_{0}%
^{\text{CR}}$ and $\phi_{0}^{\text{IP}}.$

Introducing attractive axial interactions makes CR to be stable at higher
temperatures, thus destroying IP. The higher the strength of the attractive
axial interactions, the larger $\left|  \varepsilon_{\text{a}}\right|  ,$ the
higher $T_{\text{c}}$ moves. Large enough $\left|  \varepsilon_{\text{a}%
}\right|  $ destroys the stable portion of IP resulting in second-order
liquid-liquid transition to occur in SCL region, see Fig. 5. Once
$\varepsilon$\ and $\mu$\ are kept constant, increasing $\left|
\varepsilon_{\text{p}}\right|  $ results in higher melting and Kauzmann
temperatures and larger free volume densities.

\subsection{Ordered Phases CR and IP in the Repulsive Case\ ($\varepsilon
_{\text{p}}>0)$}

\subsubsection{Absence of Solvent}

The staggered ground state (Fig. 1b) with all dimers aligned in one direction,
taken to be horizontal in Fig. 1b, is achieved only when, in addition to the
repulsive parallel-dimer interaction, an attractive axial interaction is
introduced. Moreover, the staggered state requires a different description
than the one used to describe the columnar phase studied above. This is clear
from the fact that the ground state on the cactus (consider the checker-board
version of Fig. 1b) either contains either A squares or B squares. The two
intermediate sites in a given square on the cactus are in different states
(the bond goes up or to down), but the upper and the base lattice sites are in
the same state. In order to capture this particular property of the ground
state, we need to distinguish the four sites by their color index $\alpha=$
$\mathcal{U}$,$\mathcal{D}$,$\mathcal{R}$,$\mathcal{L}$. As in the previous
case, three types of FP solutions of different symmetries are obtained: the
disordered liquid EL discussed earlier, the crystal (CR), and the intermediate
phase (IP). The situation is depicted in the Figs. 6a,b. The solid curves
represent the equilibrium states EL and CR, with a discontinuous transition
between them at the melting point $T_{\text{M}}$. We also show their
metastable extensions depicted by dashed (SCL) and dash-dot-dot (superheated
CR) curves. A liquid-liquid continuous transition occurs at $T_{\text{LL}}$
between SCL and IP. The later is shown by the dash-dot curve, which originates
at $T_{\text{LL}}$ and continues all the way down to absolute zero, just like
SCL. It is clear from Fig. 6b that both SCL and IP exhibit the entropy crisis
at $T_{\text{K}}$ and $T_{\text{K}}^{^{\prime}}.$ We show their extensions
below their respective Kauzmann temperatures by dotted curves.

The equilibrium state EL at high temperatures has been described above where
the index $\alpha$\ was omitted. Our numerical solution for the system of
1-cycle RR's that result when $\alpha$\ is taken into account predicts the
same free energy and densities as given by equations (30). However the new
solutions have a slightly different symmetry ($x_{i}^{(\mathcal{U})}%
=x_{i}^{(\mathcal{D})}$)$,$ ($x_{i}^{(\mathcal{L})}=x_{i}^{(\mathcal{R})}$)
whereas the earlier EL solution requires the symmetry ($x_{i}^{(\mathcal{U}%
)}=x_{i}^{(\mathcal{D})})=(x_{i}^{(\mathcal{L})}=x_{i}^{(\mathcal{R})}$). (The
latter symmetry can be obtained by an appropriate choice of the intitial
guesses for the new RR's.) In addition the solution also has the following
symmetry: $h^{(\alpha)}\equiv$ $h_{\text{u}}^{(\alpha)}+$ $h_{\text{d}%
}^{(\alpha)}$ $=$ $v^{(\alpha)}\equiv$ $v_{\text{u}}^{(\alpha)}+$
$v_{\text{d}}^{(\alpha)}$. The symmetry of IP is somewhat similar to that of
EL. It is also obtained in 1-cycle scheme and has the partial symmetry
$x_{i}^{(\mathcal{U})}=x_{i}^{(\mathcal{D})},$ $x_{i}^{(\mathcal{L})}%
=x_{i}^{(\mathcal{R})}$, but $h^{(\alpha)}\neq$ $v^{(\alpha)}$.

There four possible 1-cycle solutions representing four disconnected CR states
at $T=0$ are the following.
\begin{align}
\{h_{\text{d}}^{(\mathcal{R}\text{)}}  &  =h_{\text{u}}^{(\mathcal{L}\text{)}%
}=h_{\text{d}}^{(\mathcal{D}\text{)}}=h_{\text{u}}^{(\mathcal{U}\text{)}%
}=1\}\text{ or}\label{8}\\
\{h_{\text{u}}^{(\mathcal{R}\text{)}}  &  =h_{\text{d}}^{(\mathcal{L}\text{)}%
}=h_{\text{u}}^{(\mathcal{D}\text{)}}=h_{\text{d}}^{(\mathcal{U}\text{)}%
}=1\}\text{ or}\nonumber\\
\{v_{\text{d}}^{(\mathcal{R}\text{)}}  &  =v_{\text{u}}^{(\mathcal{L}\text{)}%
}=v_{\text{u}}^{(\mathcal{D}\text{)}}=v_{\text{d}}^{(\mathcal{U}\text{)}%
}=1\}\text{ or}\nonumber\\
\{v_{\text{u}}^{(\mathcal{R}\text{)}}  &  =v_{\text{d}}^{(\mathcal{L}\text{)}%
}=v_{\text{d}}^{(\mathcal{D}\text{)}}=v_{\text{u}}^{(\mathcal{U}\text{)}%
}=1\}.\nonumber
\end{align}
Only one out of the above four CR states will occur due to symmetry breaking.

The staggered phase is an example of a frozen state, which does not allow
creating any local imperfections when voids are forbidden. Once trapped in it
at low $T$, the system remains frozen in a single microstate as the
temperature is elevated further. This is signified by the
temperature-independent osmotic pressure; hence, the solution for CR is
temperature-independent. Consequently, the CR entropy $S_{\text{ord}}=0$.\ In
addition, $\phi_{\text{a}}^{\text{ord}}=1/2$ and $\phi_{\text{p}}^{\text{ord}%
}=0$ remain constant. As expected, CR is the equilibrium state at low
temperatures, while EL is the equilibrium state at high temperatures. The
melting transition between these two states takes place at $T_{\text{M}}%
\cong3.6$ where densities change discontinuously. Both CR and EL have their
extensions into the metastable region shown by means of dashed curves. The EL
extension below $T_{\text{M}}$ yields SCL. Another metastable liquid phase IP
is found in the SCL region where it meets continuously with SCL\ via a
second-order liquid-liquid transition at $T_{\text{LL}}\cong2.5.$ The new
phase IP has much higher $\phi_{\text{a}}$ and lower $\phi_{\text{p}}$
relative to SCL so that the dimers are preferentially oriented in one
direction in IP. The specific heat curves for SCL and IP show relatively large
discontinuity at $T_{\text{LL}}$.

Both SCL and IP exhibit their own entropy crises at different temperatures,
$T_{\text{K}}\cong1.1$ and $T_{\text{K}}^{^{\prime}}\cong1.7$ respectively.
Although they are not absolutely stable compared to CR, the free energy
$\omega^{\text{IP}}$ of the IP is higher than that of SCL. The calculation
predicts negative configurational entropies for the extensions of SCL and IP
below their Kauzmann temperatures represented by dotted curves in Fig. 6b. The
density values for the extensions of EL and IP reach the crystalline values at
$T=0$ so that SCL, IP, and CR have the same energy $E_{0}=\varepsilon
_{\text{a}}/2;$ see G4 above. Thus, our exact calculations confirms the
theorem of identical energies of all possible stationary states proven in
[12]. We also find that the entropies of SCL and IP are extrapolated to
$-\ln2$ and $-\ln\sqrt{2}$ respectively at $T=0$, so that $TS\rightarrow0$ as
$T\rightarrow0$, a condition for the identical energy theorem.

Changing $\varepsilon_{\text{p}}$ affects $T_{\text{M}}$ much stronger than
$T_{\text{LL}},$ vis. $T_{\text{M}}$ decreases with reduction of
$\varepsilon_{\text{p}}$.\ Figure 7 demonstrates the situation when
$T_{\text{LL}}\cong2.6$ is moved above $T_{\text{M}}\cong2.4$. We also have
$T_{\text{K}}\cong0.83$ and $T_{\text{K}}^{\prime}\cong1.1$.

The liquid-liquid transition temperature moves towards the melting temperature
as the strength $\left|  \varepsilon_{\text{a}}\right|  $ of the attractive
axial interactions increases. At the same time, the melting temperature rises
with $\left|  \varepsilon_{\text{a}}\right|  $. Reducing axial attraction
eventually makes IP to appear below $T_{\text{K}}$ only. Thus, IP is induced
by axial interactions. In the limit when $\left|  \varepsilon_{\text{a}%
}\right|  \rightarrow0,$ IP disappears from the phase diagram completely, see
Fig. 8. The inset shows that we capture the crystalline state that has
$\phi_{\text{a}}^{\text{ord}}=1/2$ at $T=0$, while $\phi_{\text{a}%
}^{\text{SCL}}$ approaches to $1/6$ at $T=0$.

The differences between the three phases in the repulsive case can be
summarized as below. The calculation shows that

\begin{itemize}
\item [(1)]EL has the property $h=v$ at each level for each direction $\alpha
$, and individual $x_{i}^{(\mathcal{U})}=x_{i}^{(\mathcal{D})},$ and
$x_{i}^{(\mathcal{L})}=x_{i}^{(\mathcal{R})}$.

\item[(2)] IP does not have the property h=v at each level for each direction
$\alpha$, but still has $x_{i}^{(\mathcal{U})}=x_{i}^{(\mathcal{D})},$ and
$x_{i}^{(\mathcal{L})}=x_{i}^{(\mathcal{R})}$.

\item[(3)] CR has neither of the properties of EL.
\end{itemize}

\subsubsection{Presence of Solvent}

The crystalline state is no longer frozen when the solvent is introduced. As
we increase $\mu$ and decrease $\varepsilon,$ the discontinuities in densities
at $T_{\text{M}}$ become smaller and smaller. Eventually melting transforms
into a second-order transition. The presence of the solvent makes
$T_{\text{M}}$ go down; The larger $\mu$ and the smaller $\varepsilon$ are,
the smaller $T_{\text{M}}$ becomes. Figure 9 shows the solvent density
$\phi_{0}$. Compared to the case of $\varepsilon_{\text{p}}<0,$ the values of
$\phi_{0}$ at the Kauzmann temperatures are much larger. The solvent density
in SCL is larger than in IP and the same holds for the ratio $D_{0}.$ From
Fig. 9, we immediately note that the relative amount of dimerized solvents is
higher in SCL than in IP for this case.

\section{Discussion and Conclusions}

We have argued that the configurational entropy is the central concept and has
to be defined properly if we have any chance of success in explaining the
experimentally observed glass transition in supercooled viscous liquids. We
have followed the conventional definition of the configurational entropy,
which is obtained by considering the configurational degrees of freedom and is
obtained by subtracting the entropy $S_{\text{KE}}(T)$ due to the
translational degrees of freedom from the total entropy. It is this
configurational entropy that is used to define the configurational partition
function. This definition is compared with other definitions of
configurational entropy used by experimentalists and have shown that their
definitions are operational in spirit to estimate the configurational entropy
since the latter is not experimentally accessible, at least at present by any
known technique.

Following the original suggestion of Kauzmann, we have adopted the view that
observing $S_{\text{SCL}}<0$ under extrapolation of experimental values of
$S_{\text{SCL}}$ implies that such extrapolated states cannot exist. Any state
to be observed in Nature must, as a prerequisite, have a non-negative entropy.
This principle must be obeyed by any observed state in Nature, regardless of
whether the state is an equilibrium state, a strongly time-dependent
metastable state, or a stationary metastable state. We have shown that while
it is possible in theoretical investigations to probe the stationary limit of
SCL or other metastable states and be able to investigate the ideal glass
transition, which is invoked to avoid the configurational entropy crisis
($S_{\text{SCL}}<0$), it is impossible to establish the existence of the ideal
glass transition without the help of extrapolation in experiments or numerical
simulation, since they only deal with states that exist in Nature.
Consequently, they will only access states for which the configurational
entropy must strictly be non-negative.

Since the value of the configurational entropy is crucial in locating the
Kauzmann temperature, we have shown how to define the entropy properly
($S\geq0$) by discretizing the real and momentum spaces. This procedure
ensures $S\geq0.$ We then argue that since the translational degrees of
freedom are decoupled from the configurational degrees of freedom, we need to
ensure the non-negativity of each of the entropy from the two degrees of
freedom. Since the entropy due to the kinetic energy is always non-negative,
we need to only verify whether the configurational entropy is non-negative so
that the entropy crisis would not occur.

One of the major aims of the work was to establish the existence of the
entropy crisis in metastable molecular fluids. We have developed a simple
model of molecular fluids composed of dimers. The model is defined on a
lattice so that it only contains configurational entropy and contains
directional and direction-independent interactions. One of the former
interactions is the interaction between the end-points of the dimers, which is
either attractive or repulsive, and is used to broadly classify the molecular
liquid into attractive and repulsive cases. This is because the CR-symmetry in
the two cases are very different, as shown in Fig. 1.

The model is solved exactly on a special kind of recursive lattices, commonly
known as the site-sharing Husimi cactus. The cactus is an approximation of a
square lattice and shares with it the coordination number and the smallest
loop size. The method of solution uses recursive technique, which is standard
by now. What distinguishes the present analysis with most other similar
investigations is that the ordered phase (CR\ and IP) descriptions require
identifying novel FP structures with appropriate symmetries, which are very
different from the symmetry of the FP solution describing the disordered phase
(EL and its metastable extension SCL). This also requires us to calculate the
osmotic pressure ($\eta>0$) or the free energy ($\eta=0$) appropriately.

Because of the directional interactions, we also obtain an intermediate phase
IP, which may or may not be an equilibrium state depending on the directional
interactions. In the attractive case, IP is an equilibrium state, but its
extension at low temperatures is a metastable state with respect to CR.
However, for the repulsive case, it is a metastable state in Fig. 5 and an
equilibrium state in Fig. 6. In both cases, its extension at low temperatures
is a metastable state with respect to CR. Both EL and IP have their extensions
that exhibit their own entropy crisis. The fact that EL's metastable extension
SCL exhibits an entropy crisis is not a surprise on its own, since this is
what we intended to demonstrate, it came as a surprise that the metastable
extension of IP (which itself may be a metastable state with respect to CR,
but an ''equilibrium state'' with respect to SCL) also exhibited a Kauzmann
temperature of its own. This is consistent with the rigorous analysis of
Gujrati [12], which demonstrates the existence of an entropy crisis in a state
which is a metastable extension with respect to CR. The argument is applicable
to any number of metastable extensions, as long as these extensions are of
stationary nature.

The model is rich enough due to its complex energetics that we also find
liquid-liquid transition in the model. The transition is driven by the
directional interactions in the model. The distributions of solvent particles
in the metastable states are very different from that in CR. This is seen
clearly from the behavior of $D_{0}.$

The analysis of the present work also confirms the equal energy principle G3
at $T=0$ for all states that can be continued to $T=0$; here, we are carrying
out the continuation without any regard to the principle of reality ($S\geq
0$). This is important as it immediately implies that there must be a positive
Kauzmann temperature. We further note that $TS(T)\rightarrow0$ for all of the
states. This condition is a prerequisite for the Gujrati proof [12].

Because of \ the equal energy observation, the question that naturally arises
is whether all the states at absolute zeo represent the same state. The answer
is negative in general, as we demonstrate now. It is clear that all densities
that determine the energy must be the same in all states, since the equal
energy principle is obeyed no matter what the values of the interaction energy
parameters. However, there are other geometrical and topological quantities in
the system that do not affect the energy. These quantities have no reasion to
be the same in all states at $T=0.$ We have considered the repulsive case in
which we let $\varepsilon_{\text{a}}=0$, so that $\phi_{\text{a}}$ does not
affect the energy of the system; see Fig. 8. We find that $\phi_{\text{a}%
}^{\text{CR}}=1/2,$ while $\phi_{\text{a}}^{\text{SCL}}=1/6,$ so that it does
not have the same value in all states at $T=0$. Thus, the states at absolute
zero are different despite having the same energy.

In conjunction with the earlier results demonstrating the existence of a
positive Kauzmann temperature for long polymers, the present work fills the
gap by showing that even molecular viscous fluids also have a positive
Kauzmann temperature. We thus conclude that the stationary metastable state
must always exhibit a positive Kauzmann temperature below which the
configurational entropy will become negative no matter what the molecular size
is, if we insist on extrapolation. Since a negative entropy violates the
principle of reality, we are forced to conclude that a new state, the ideal
glass state, must be brought into the picture to replace the extrapolated
state below the Kauzmann temperature. This is, again, in conformity with G4.

Thus, we finally conclude that every (stationary) metastable state must
experience an entropy crisis at a non-negative Kauzmann temperature, no matter
what the size of the molecules are. The free volume does not play a
determining role, except when it becomes too large to destroy the entropy
crisis by bringing down the Kauzmann temperature to absolute zero. Our
analysis says nothing about systems that have no ordered state for which,
therefore, stationary metastability is not an issue.

\section{Appendix}%

\begin{align*}
P_{\text{s}}^{(\mathcal{U})}  &  =\eta w^{2}(w_{\text{a}}w_{\text{p}%
}h_{\text{u}}^{(\mathcal{L})}h_{\text{u}}^{(\mathcal{R})}h_{\text{u}%
}^{(\mathcal{U})}+h_{\text{u}}^{(\mathcal{L})}h_{\text{u}}^{(\mathcal{R}%
)}v_{\text{u}}^{(\mathcal{U})}+w_{\text{p}}v_{\text{u}}^{(\mathcal{L}%
)}h_{\text{u}}^{(\mathcal{R})}v_{\text{u}}^{(\mathcal{U})}\\
&  +w_{\text{p}}v_{\text{u}}^{(\mathcal{L})}h_{\text{u}}^{(\mathcal{R}%
)}h_{\text{u}}^{(\mathcal{U})}+w_{\text{a}}h_{\text{u}}^{(\mathcal{L}%
)}v_{\text{u}}^{(\mathcal{R})}h_{\text{u}}^{(\mathcal{U})}+w_{\text{a}%
}h_{\text{u}}^{(\mathcal{L})}v_{\text{u}}^{(\mathcal{R})}v_{\text{u}%
}^{(\mathcal{U})}\\
&  +w_{\text{a}}w_{\text{p}}v_{\text{u}}^{(\mathcal{L})}v_{\text{u}%
}^{(\mathcal{R})}v_{\text{u}}^{(\mathcal{U})}+v_{\text{u}}^{(\mathcal{L}%
)}v_{\text{u}}^{(\mathcal{R})}h_{\text{u}}^{(\mathcal{U})}+h_{\text{d}%
}^{(\mathcal{L})}h_{\text{u}}^{(\mathcal{R})}h_{\text{d}}^{(\mathcal{U})}\\
&  +h_{\text{d}}^{(\mathcal{L})}v_{\text{u}}^{(\mathcal{R})}h_{\text{d}%
}^{(\mathcal{U})}+h_{\text{u}}^{(\mathcal{L})}v_{\text{d}}^{(\mathcal{R}%
)}v_{\text{d}}^{(\mathcal{U})}+v_{\text{u}}^{(\mathcal{L})}v_{\text{d}%
}^{(\mathcal{R})}v_{\text{d}}^{(\mathcal{U})}\\
&  +w_{\text{p}}s^{(\mathcal{L})}h_{\text{u}}^{(\mathcal{R})}h_{\text{u}%
}^{(\mathcal{U})}+s^{(\mathcal{L})}h_{\text{u}}^{(\mathcal{R})}v_{\text{u}%
}^{(\mathcal{U})}+s^{(\mathcal{L})}v_{\text{u}}^{(\mathcal{R})}h_{\text{u}%
}^{(\mathcal{U})}\\
&  +w_{\text{a}}s^{(\mathcal{L})}v_{\text{u}}^{(\mathcal{R})}v_{\text{u}%
}^{(\mathcal{U})}+w^{2}h_{\text{u}}^{(\mathcal{L})}h_{\text{u}}^{(\mathcal{R}%
)}s^{(\mathcal{U})}+w^{2}v_{\text{u}}^{(\mathcal{L})}h_{\text{u}%
}^{(\mathcal{R})}s^{(\mathcal{U})}\\
&  +w^{2}h_{\text{u}}^{(\mathcal{L})}v_{\text{u}}^{(\mathcal{R})}%
s^{(\mathcal{U})}+w^{2}v_{\text{u}}^{(\mathcal{L})}v_{\text{u}}^{(\mathcal{R}%
)}s^{(\mathcal{U})}+h_{\text{u}}^{(\mathcal{L})}s^{(\mathcal{R})}v_{\text{u}%
}^{(\mathcal{U})}\\
&  +w_{\text{a}}h_{\text{u}}^{(\mathcal{L})}s^{(\mathcal{R})}h_{\text{u}%
}^{(\mathcal{U})}+w_{\text{p}}v_{\text{u}}^{(\mathcal{L})}s^{(\mathcal{R}%
)}v_{\text{u}}^{(\mathcal{U})}+v_{\text{u}}^{(\mathcal{L})}s^{(\mathcal{R}%
)}h_{\text{u}}^{(\mathcal{U})}\\
&  +h_{\text{u}}^{(\mathcal{L})}s^{(\mathcal{R})}s^{(\mathcal{U})}%
+v_{\text{u}}^{(\mathcal{L})}s^{(\mathcal{R})}s^{(\mathcal{U})}%
+s^{(\mathcal{L})}s^{(\mathcal{R})}v_{\text{u}}^{(\mathcal{U})}\\
&  +s^{(\mathcal{L})}s^{(\mathcal{R})}h_{\text{u}}^{(\mathcal{U}%
)}+s^{(\mathcal{L})}h_{\text{u}}^{(\mathcal{R})}s^{(\mathcal{U})}%
+s^{(\mathcal{L})}v_{\text{u}}^{(\mathcal{R})}s^{(\mathcal{U})}\\
&  +h_{\text{d}}^{(\mathcal{L})}s^{(\mathcal{R})}h_{\text{d}}^{(\mathcal{U}%
)}+s^{(\mathcal{L})}v_{\text{d}}^{(\mathcal{R})}v_{\text{d}}^{(\mathcal{U}%
)})+\eta s^{(\mathcal{L})}s^{(\mathcal{R})}s^{(\mathcal{U})},
\end{align*}%
\begin{align*}
P_{\text{h}_{\text{d}}}^{(\mathcal{U})}  &  =w_{\text{a}}^{2}w_{\text{p}}%
^{2}h_{\text{u}}^{(\mathcal{L})}h_{\text{u}}^{(\mathcal{R})}h_{\text{u}%
}^{(\mathcal{U})}+w_{\text{a}}w_{\text{p}}h_{\text{u}}^{(\mathcal{L}%
)}h_{\text{u}}^{(\mathcal{R})}v_{\text{u}}^{(\mathcal{U})}+w_{\text{a}%
}w_{\text{p}}v_{\text{u}}^{(\mathcal{L})}h_{\text{u}}^{(\mathcal{R})}%
v_{u}^{(\mathcal{U})}\\
&  +w_{\text{a}}w_{\text{p}}v_{\text{u}}^{(\mathcal{L})}h_{\text{u}%
}^{(\mathcal{R})}h_{\text{u}}^{(\mathcal{U})}+w_{\text{a}}w_{\text{p}%
}h_{\text{u}}^{(\mathcal{L})}v_{\text{u}}^{(\mathcal{R})}h_{\text{u}%
}^{(\mathcal{U})}+w_{\text{a}}w_{\text{p}}h_{\text{u}}^{(\mathcal{L}%
)}v_{\text{u}}^{(\mathcal{R})}v_{\text{u}}^{(\mathcal{U})}\\
&  +w_{a}w_{p}v_{\text{u}}^{(\mathcal{L})}v_{\text{u}}^{(\mathcal{R}%
)}v_{\text{u}}^{(\mathcal{U})}+v_{\text{u}}^{(\mathcal{L})}v_{\text{u}%
}^{(\mathcal{R})}h_{\text{u}}^{(\mathcal{U})}+w_{\text{a}}h_{\text{d}%
}^{(\mathcal{L})}h_{\text{u}}^{(\mathcal{R})}h_{\text{d}}^{(\mathcal{U})}\\
&  +h_{\text{d}}^{(\mathcal{L})}v_{\text{u}}^{(\mathcal{R})}h_{\text{d}%
}^{(\mathcal{U})}+w_{\text{p}}h_{\text{u}}^{(\mathcal{L})}v_{\text{d}%
}^{(\mathcal{R})}v_{\text{d}}^{(\mathcal{U})}+v_{\text{u}}^{(\mathcal{L}%
)}v_{\text{d}}^{(\mathcal{R})}v_{\text{d}}^{(\mathcal{U})}\\
&  +w^{2}(w_{\text{a}}w_{\text{p}}s^{(\mathcal{L})}h_{\text{u}}^{(\mathcal{R}%
)}h_{\text{u}}^{(\mathcal{U})}+w_{\text{a}}s^{(\mathcal{L})}h_{\text{u}%
}^{(\mathcal{R})}v_{\text{u}}^{(\mathcal{U})}+s^{(\mathcal{L})}v_{\text{u}%
}^{(\mathcal{R})}h_{\text{u}}^{(\mathcal{U})}\\
&  +w_{\text{a}}s^{(\mathcal{L})}v_{\text{u}}^{(\mathcal{R})}v_{\text{u}%
}^{(\mathcal{U})}+w_{\text{a}}w_{\text{p}}h_{\text{u}}^{(\mathcal{L}%
)}h_{\text{u}}^{(\mathcal{R})}s^{(\mathcal{U})}+w_{\text{a}}v_{\text{u}%
}^{(\mathcal{L})}h_{\text{u}}^{(\mathcal{R})}s^{(\mathcal{U})}\\
&  +w_{\text{p}}h_{\text{u}}^{(\mathcal{L})}v_{\text{u}}^{(\mathcal{R}%
)}s^{(\mathcal{U})}+v_{\text{u}}^{(\mathcal{L})}v_{\text{u}}^{(\mathcal{R}%
)}s^{(\mathcal{U})}+w_{\text{p}}h_{\text{u}}^{(\mathcal{L})}s^{(\mathcal{R}%
)}v_{\text{u}}^{(\mathcal{U})}\\
&  +w_{\text{a}}w_{\text{p}}h_{\text{u}}^{(\mathcal{L})}s^{(\mathcal{R}%
)}h_{\text{u}}^{(\mathcal{U})}+w_{\text{p}}v_{\text{u}}^{(\mathcal{L}%
)}s^{(\mathcal{R})}v_{\text{u}}^{(\mathcal{U})}+v_{\text{u}}^{(\mathcal{L}%
)}s^{(\mathcal{R})}h_{\text{u}}^{(\mathcal{U})}\\
&  +w_{\text{a}}s^{(\mathcal{L})}h_{\text{u}}^{(\mathcal{R})}s^{(\mathcal{U}%
)}+s^{(\mathcal{L})}v_{\text{u}}^{(\mathcal{R})}s^{(\mathcal{U})}+h_{\text{d}%
}^{(\mathcal{L})}s^{(\mathcal{R})}h_{\text{d}}^{(\mathcal{U})}\\
&  +s^{(\mathcal{L})}v_{\text{d}}^{(\mathcal{R})}v_{\text{d}}^{(\mathcal{U}%
)}+s^{(\mathcal{L})}s^{(\mathcal{R})}s^{(\mathcal{U})}),
\end{align*}%
\begin{align*}
P_{\text{v}_{\text{d}}}^{(\mathcal{U})}  &  =w_{\text{a}}w_{\text{p}%
}h_{\text{u}}^{(\mathcal{L})}h_{\text{u}}^{(\mathcal{R})}h_{\text{u}%
}^{(\mathcal{U})}+h_{\text{u}}^{(\mathcal{L})}h_{\text{u}}^{(\mathcal{R}%
)}v_{\text{u}}^{(\mathcal{U})}+w_{\text{a}}w_{\text{p}}v_{\text{u}%
}^{(\mathcal{L})}h_{\text{u}}^{(\mathcal{R})}v_{\text{u}}^{(\mathcal{U})}\\
&  +w_{\text{a}}w_{\text{p}}v_{\text{u}}^{(\mathcal{L})}h_{\text{u}%
}^{(\mathcal{R})}h_{\text{u}}^{(\mathcal{U})}+w_{\text{a}}w_{\text{p}%
}h_{\text{u}}^{(\mathcal{L})}v_{\text{u}}^{(\mathcal{R})}h_{\text{u}%
}^{(\mathcal{U})}+w_{\text{a}}w_{\text{p}}h_{\text{u}}^{(\mathcal{L}%
)}v_{\text{u}}^{(\mathcal{R})}v_{\text{u}}^{(\mathcal{U})}\\
&  +w_{\text{a}}^{2}w_{\text{p}}^{2}v_{\text{u}}^{(\mathcal{L})}v_{\text{u}%
}^{(\mathcal{R})}v_{\text{u}}^{(\mathcal{U})}+w_{\text{a}}w_{\text{p}%
}v_{\text{u}}^{(\mathcal{L})}v_{\text{u}}^{(\mathcal{R})}h_{\text{u}%
}^{(\mathcal{U})}+h_{\text{d}}^{(\mathcal{L})}h_{\text{u}}^{(\mathcal{R}%
)}h_{\text{d}}^{(\mathcal{U})}\\
&  +w_{\text{p}}h_{\text{d}}^{(\mathcal{L})}v_{\text{u}}^{(\mathcal{R}%
)}h_{\text{d}}^{(\mathcal{U})}+h_{\text{u}}^{(\mathcal{L})}v_{\text{d}%
}^{(\mathcal{R})}v_{\text{d}}^{(\mathcal{U})}+w_{\text{a}}v_{\text{u}%
}^{(\mathcal{L})}v_{\text{d}}^{(\mathcal{R})}v_{\text{d}}^{(\mathcal{U})}\\
&  +w^{2}(w_{\text{p}}s^{(\mathcal{L})}h_{\text{u}}^{(\mathcal{R})}%
h_{\text{u}}^{(\mathcal{U})}+s^{(\mathcal{L})}h_{\text{u}}^{(\mathcal{R}%
)}v_{\text{u}}^{(\mathcal{U})}+w_{\text{p}}s^{(\mathcal{L})}v_{\text{u}%
}^{(\mathcal{R})}h_{\text{u}}^{(\mathcal{U})}\\
&  +w_{\text{a}}w_{\text{p}}s^{(\mathcal{L})}v_{\text{u}}^{(\mathcal{R}%
)}v_{\text{u}}^{(\mathcal{U})}+h_{\text{u}}^{(\mathcal{L})}h_{\text{u}%
}^{(\mathcal{R})}s^{(\mathcal{U})}+w_{\text{a}}v_{\text{u}}^{(\mathcal{L}%
)}h_{\text{u}}^{(\mathcal{R})}s^{(\mathcal{U})}\\
&  +w_{\text{p}}h_{\text{u}}^{(\mathcal{L})}v_{\text{u}}^{(\mathcal{R}%
)}s^{(\mathcal{U})}+w_{\text{a}}w_{\text{p}}v_{\text{u}}^{(\mathcal{L}%
)}v_{\text{u}}^{(\mathcal{R})}s^{(\mathcal{U})}+h_{\text{u}}^{(\mathcal{L}%
)}s^{(\mathcal{R})}v_{\text{u}}^{(\mathcal{U})}\\
&  +w_{\text{a}}h_{\text{u}}^{(\mathcal{L})}s^{(\mathcal{R})}h_{\text{u}%
}^{(\mathcal{U})}+w_{\text{p}}w_{\text{a}}v_{\text{u}}^{(\mathcal{L}%
)}s^{(\mathcal{R})}v_{\text{u}}^{(\mathcal{U})}+w_{\text{a}}v_{\text{u}%
}^{(\mathcal{L})}s^{(\mathcal{R})}h_{\text{u}}^{(\mathcal{U})}\\
&  +h_{\text{u}}^{(\mathcal{L})}s^{(\mathcal{R})}s^{(\mathcal{U})}%
+w_{\text{a}}v_{\text{u}}^{(\mathcal{L})}s^{(\mathcal{R})}s^{(\mathcal{U}%
)}+w^{2}s^{(\mathcal{L})}s^{(\mathcal{R})}v_{\text{u}}^{(\mathcal{U})}\\
&  +w^{2}s^{(\mathcal{L})}s^{(\mathcal{R})}h_{\text{u}}^{(\mathcal{U}%
)}+s^{(\mathcal{L})}h_{\text{u}}^{(\mathcal{R})}s^{(\mathcal{U})}+w_{\text{p}%
}s^{(\mathcal{L})}v_{\text{u}}^{(\mathcal{R})}s^{(\mathcal{U})}\\
&  +h_{\text{d}}^{(\mathcal{L})}s^{(\mathcal{R})}h_{\text{d}}^{(\mathcal{U}%
)}+s^{(\mathcal{L})}v_{\text{d}}^{(\mathcal{R})}v_{\text{d}}^{(\mathcal{U}%
)}+s^{(\mathcal{L})}s^{(\mathcal{R})}s^{(\mathcal{U})}),
\end{align*}%
\begin{align*}
P_{\text{v}_{\text{u}}}^{(\mathcal{U})}  &  =w_{\text{p}}v_{\text{d}%
}^{(\mathcal{L})}h_{\text{u}}^{(\mathcal{R})}h_{\text{u}}^{(\mathcal{U}%
)}+v_{\text{d}}^{(\mathcal{L})}v_{\text{u}}^{(\mathcal{R})}h_{\text{u}%
}^{(\mathcal{U})}+v_{\text{d}}^{(\mathcal{L})}h_{\text{u}}^{(\mathcal{R}%
)}v_{\text{u}}^{(\mathcal{U})}\\
&  +w_{\text{a}}v_{\text{d}}^{(\mathcal{L})}v_{\text{u}}^{(\mathcal{R}%
)}v_{\text{u}}^{(\mathcal{U})}+w_{\text{p}}^{2}v_{\text{d}}^{(\mathcal{L}%
)}v_{\text{d}}^{(\mathcal{R})}v_{\text{d}}^{(\mathcal{U})}+w^{2}(v_{\text{d}%
}^{(\mathcal{L})}s^{(\mathcal{R})}v_{\text{u}}^{(\mathcal{U})}\\
&  +v_{\text{d}}^{(\mathcal{L})}s^{(\mathcal{R})}h_{\text{u}}^{(\mathcal{U}%
)}+v_{\text{d}}^{(\mathcal{L})}h_{\text{u}}^{(\mathcal{R})}s^{(\mathcal{U}%
)}+v_{\text{d}}^{(\mathcal{L})}v_{\text{u}}^{(\mathcal{R})}s^{(\mathcal{U}%
)}+v_{\text{d}}^{(\mathcal{L})}s^{(\mathcal{R})}s^{(\mathcal{U})}),
\end{align*}%
\begin{align*}
P_{\text{h}_{\text{u}}}^{(\mathcal{U})}  &  =w_{\text{p}}v_{\text{u}%
}^{(\mathcal{L})}h_{\text{d}}^{(\mathcal{R})}v_{\text{u}}^{(\mathcal{U}%
)}+h_{\text{u}}^{(\mathcal{L})}h_{\text{d}}^{(\mathcal{R})}v_{\text{u}%
}^{(\mathcal{U})}+v_{\text{u}}^{(\mathcal{L})}h_{\text{d}}^{(\mathcal{R}%
)}h_{\text{u}}^{(\mathcal{U})}\\
&  +w_{\text{a}}h_{\text{u}}^{(\mathcal{L})}h_{\text{d}}^{(\mathcal{R}%
)}h_{\text{u}}^{(\mathcal{U})}+w_{\text{p}}^{2}h_{\text{d}}^{(\mathcal{L}%
)}h_{\text{d}}^{(\mathcal{R})}h_{\text{d}}^{(\mathcal{U})}+w^{2}%
(s^{(\mathcal{L})}h_{\text{d}}^{(\mathcal{R})}v_{\text{u}}^{(\mathcal{U})}\\
&  +s^{(\mathcal{L})}h_{\text{d}}^{(\mathcal{R})}h_{\text{u}}^{(\mathcal{U}%
)}+h_{\text{u}}^{(\mathcal{L})}h_{\text{d}}^{(\mathcal{R})}s^{(\mathcal{U}%
)}+v_{\text{u}}^{(\mathcal{L})}h_{\text{d}}^{(\mathcal{R})}s^{(\mathcal{U}%
)}+s^{(\mathcal{L})}h_{\text{d}}^{(\mathcal{R})}s^{(\mathcal{U})}),
\end{align*}

$P_{i}^{(\mathcal{D})}$ is obtained from $P_{i}^{(\mathcal{U})}$ by
interchange $x_{j}^{(\mathcal{R})}\leftrightarrow x_{j}^{(\mathcal{L})},$
$x_{k}^{(\mathcal{U})}\rightarrow x_{k}^{(\mathcal{D})}.$

$P_{i}^{(\mathcal{L})}$ is obtained from $P_{i}^{(\mathcal{U})}$ by
interchange $x_{j}^{(\mathcal{L})}\leftrightarrow x_{j}^{(\mathcal{U})},$
$x_{k}^{(\mathcal{R})}\rightarrow x_{k}^{(\mathcal{D})}.$

$P_{i}^{(\mathcal{R})}$ is obtained from $P_{i}^{(\mathcal{U})}$ by
interchange $x_{j}^{(\mathcal{R})}\leftrightarrow x_{j}^{(\mathcal{U})},$
$x_{k}^{(\mathcal{L})}\rightarrow x_{k}^{(\mathcal{D})}.$

\begin{figure}[p]
\begin{center}
\epsfig{file=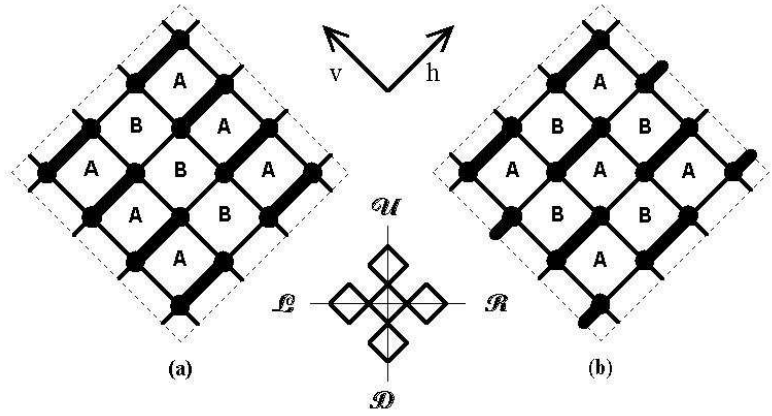,width=4.5in}
\end{center}
\caption{Columnar (a) and staggered (b) states, and the bond (h,v) and the
lattice ($\mathcal{R},\mathcal{L},\mathcal{D},\mathcal{U}$) directions.}%
\end{figure}

\begin{figure}[p]
\begin{center}
\epsfig{file=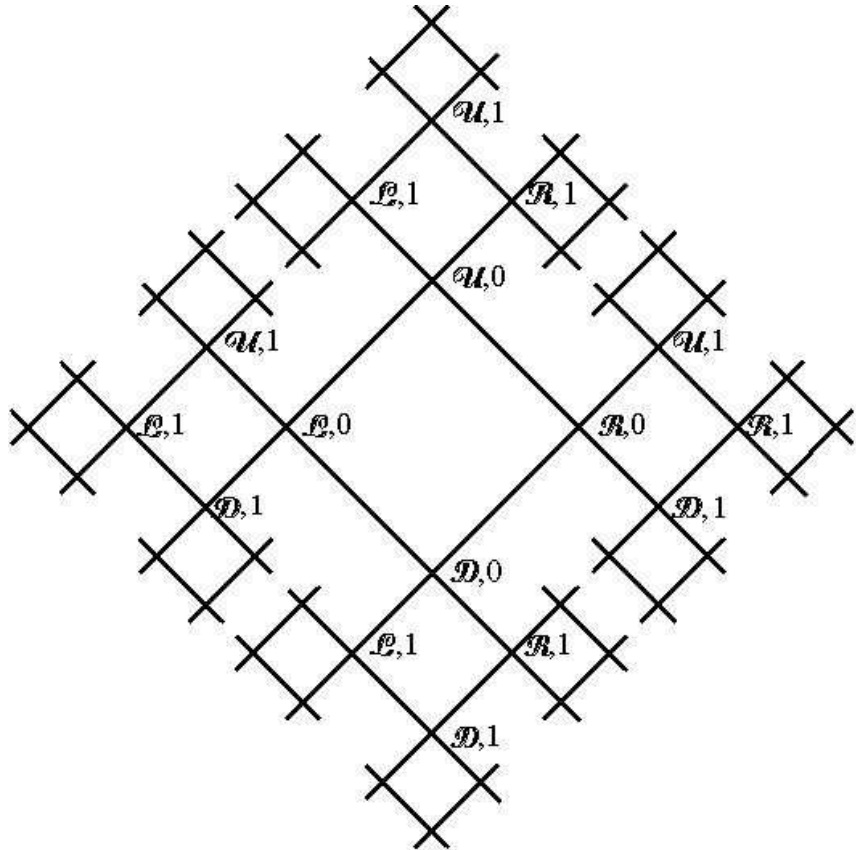,width=4.5in}
\end{center}
\caption{Hisimi tree with labels representing distance from the origin, levels
\textit{m,} and representing orientation with respect to the origin,
directional indices $\alpha= \mathcal{R},\mathcal{L},\mathcal{D},\mathcal{U}%
$.}%
\end{figure}

\begin{figure}[p]
\begin{center}
\epsfig{file=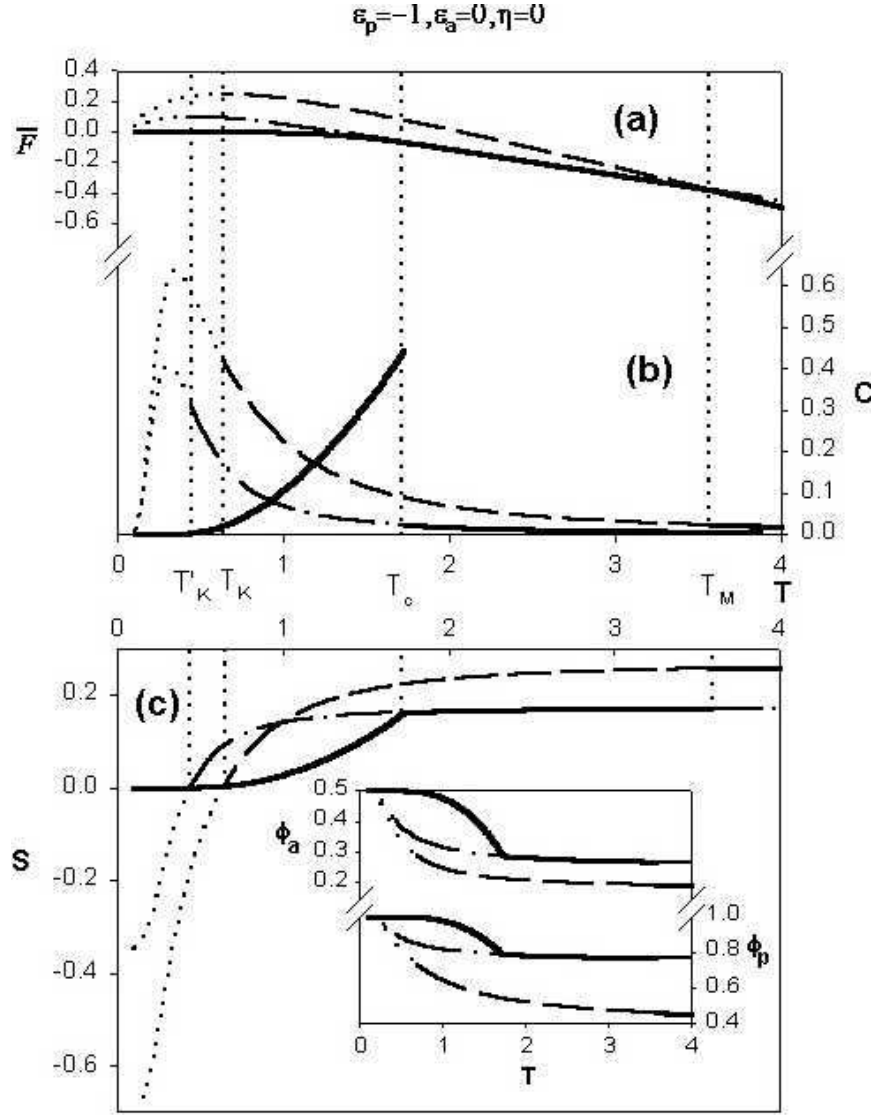,width=4.5in}
\end{center}
\caption{Phase diagram for the case of attractive parallel-dimer interaction
and no solvent; temperature dependencies of (a) the shifted thermodynamic
potential, (b) the specific heat $C$, (c) the entropy and in the inset of (c)
the contact densities $\phi_{\text{a}}$ and $\phi_{\text{p}}$.}%
\end{figure}

\begin{figure}[p]
\begin{center}
\epsfig{file=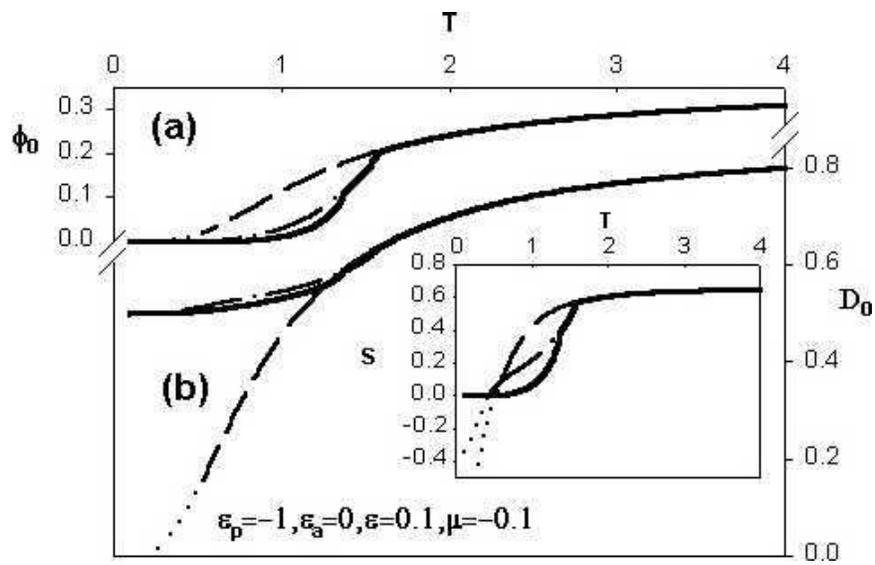,width=4.5in}
\end{center}
\caption{Illustration of the effect of free volume for $\varepsilon_{\text{p}%
}<0,\varepsilon_{\text{a}}=0$. Temperature dependencies of (a) the free
volume, (b) the ratio $D_{0}=\phi_{00}/\phi_{0}$, and the entropy in the
inset.}%
\end{figure}

\begin{figure}[p]
\begin{center}
\epsfig{file=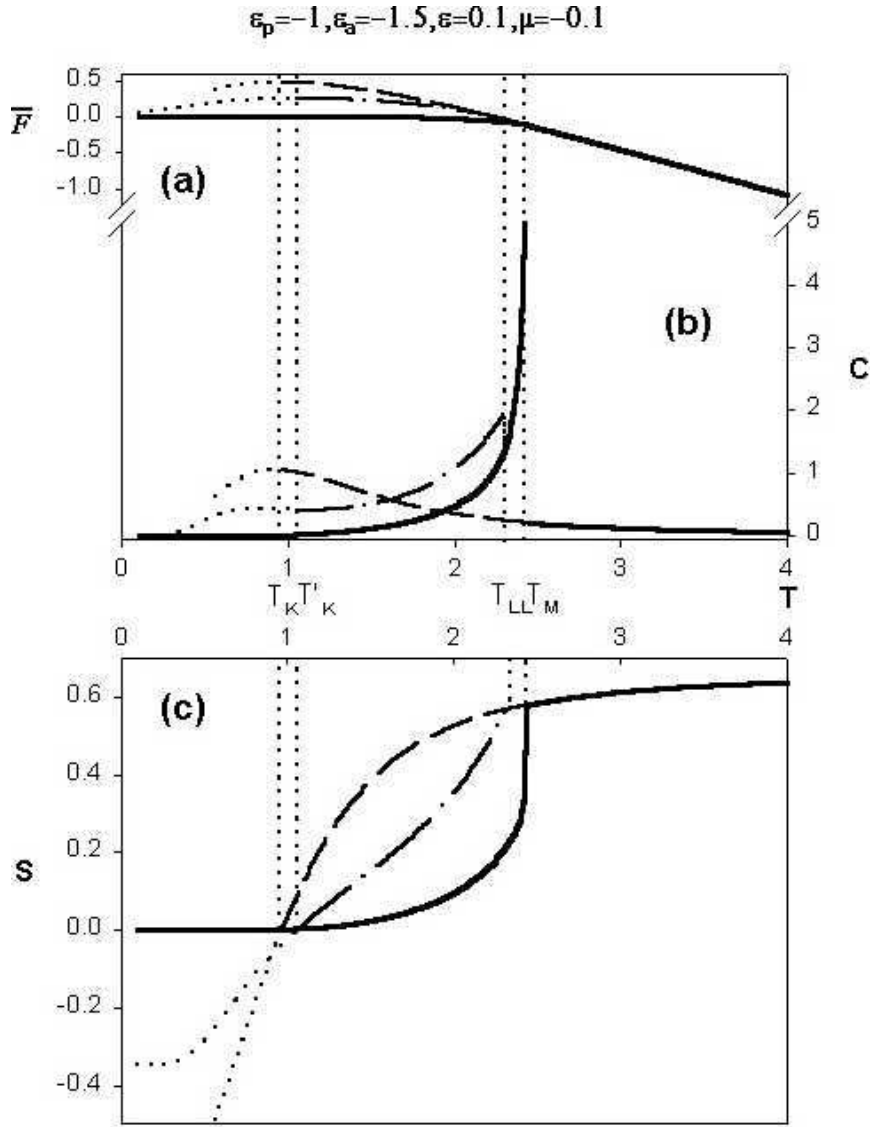,width=4.5in}
\end{center}
\caption{Plot demostrating the effect of attractive axial interaction for
$\varepsilon_{p}<0$. Increasing the strength of the attractive axial
interaction makes IP to apperar in the SCL region.}%
\end{figure}

\begin{figure}[p]
\begin{center}
\epsfig{file=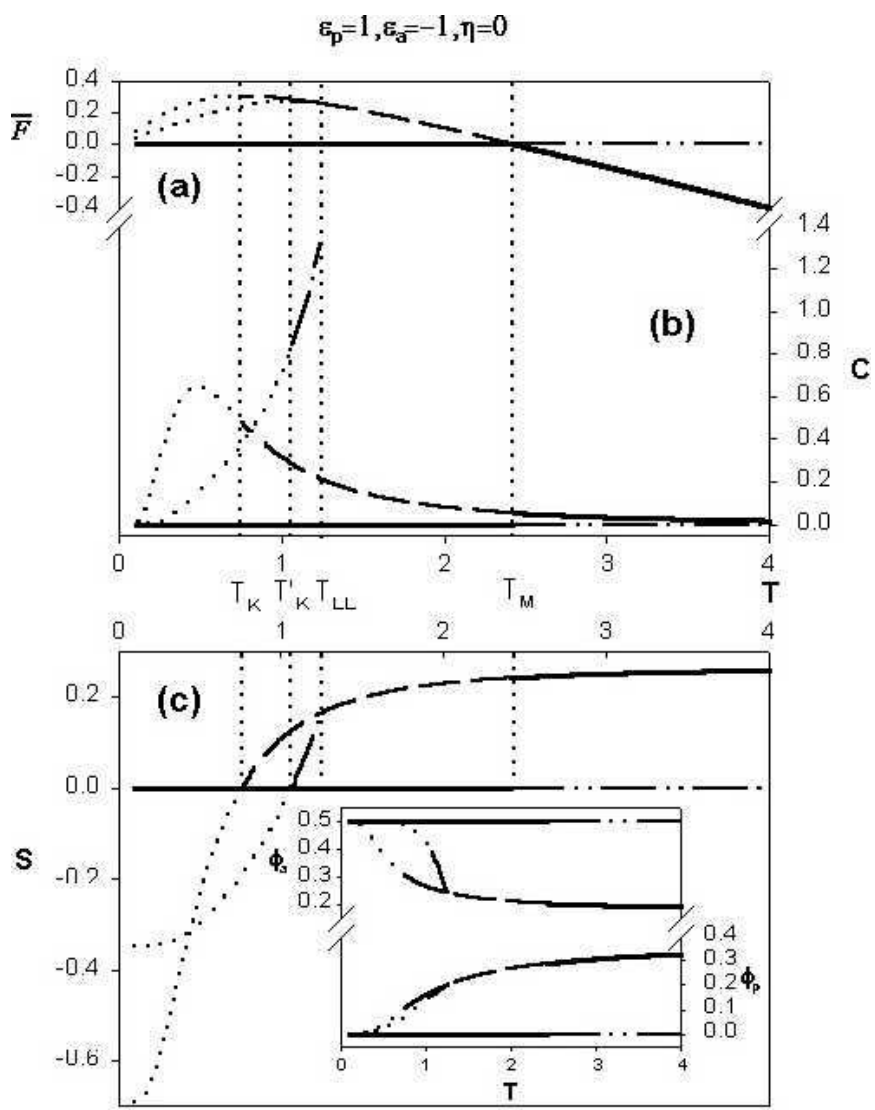,width=4.5in}
\end{center}
\caption{Phase diagram for the case of repulsive parallel-dimer interaction
and no solvent. The liquid-liquid transition at $T_{\text{LL}}$ is below the
melting point at $T_{\text{M}}$.}%
\end{figure}

\begin{figure}[p]
\begin{center}
\epsfig{file=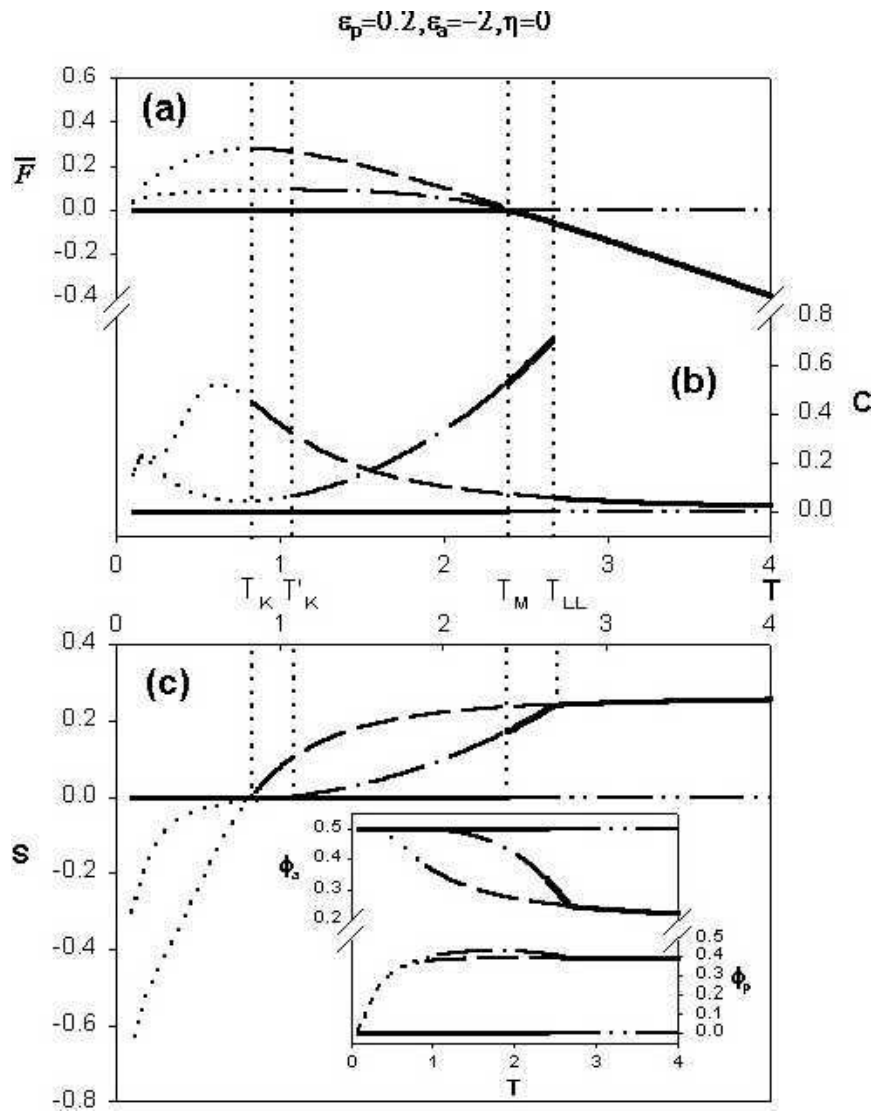,width=4.5in}
\end{center}
\caption{Phase for the repulsive parallel-dimer interaction representing the
liquid-liquid transition above the melting point.}%
\end{figure}

\begin{figure}[p]
\begin{center}
\epsfig{file=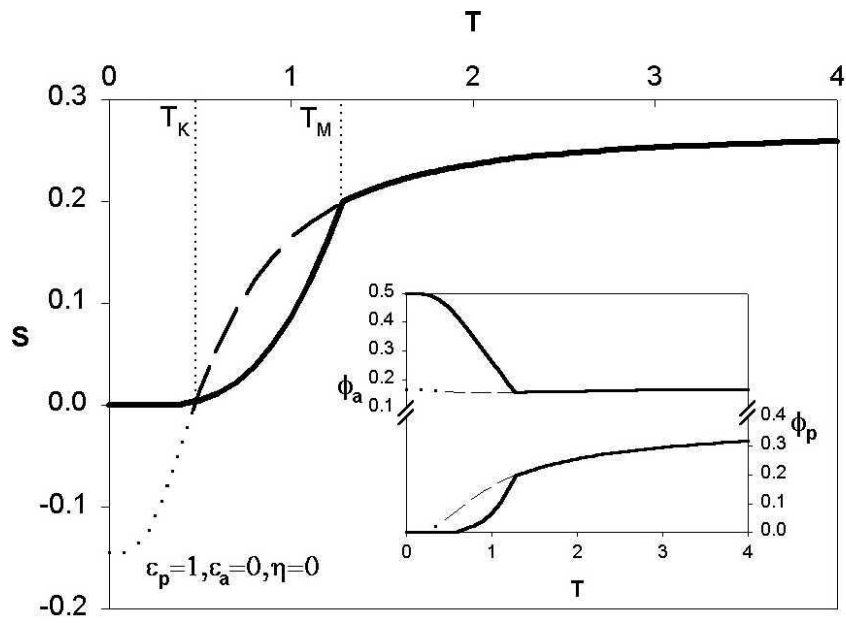,width=4.5in}
\end{center}
\caption{Phase diagram for the case as the strength of the axial interaction is
tending to zero. }%
\end{figure}

\begin{figure}[p]
\begin{center}
\epsfig{file=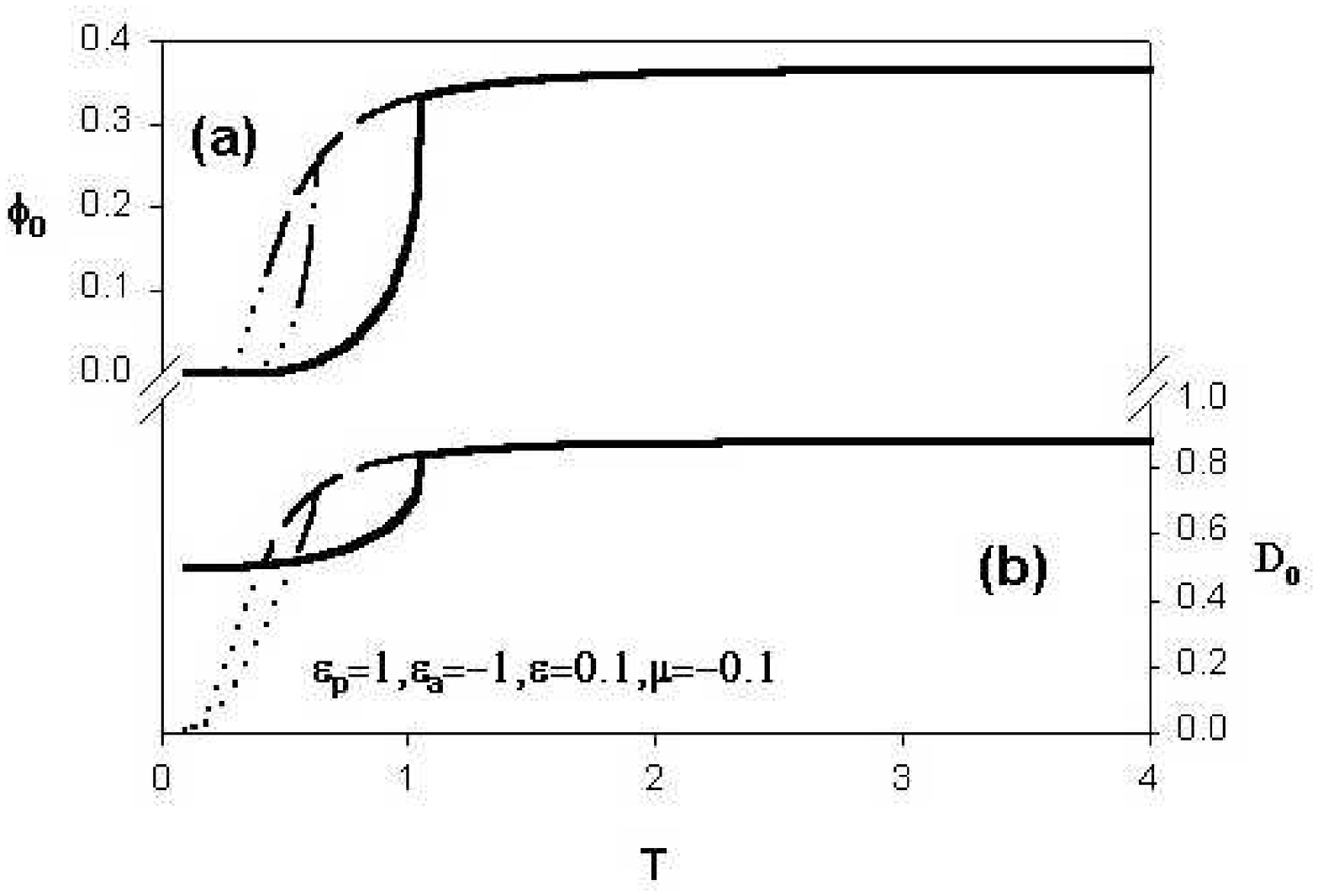,width=4.5in}
\end{center}
\caption{Temperature dependencies of (a) the free volume and (b) the ratio
$D_{0}=\phi_{00}/\phi_{0}$ for the case of repulsive parallel-dimer
interaction.}%
\end{figure}


\begin{thebibliography}{99}
\bibitem{note0}For a microscopic system, the contribution due to $\ln dE$ (or
due to $\ln dP;$ see later) to the entropy is negligible, and will be
neglected in the following. Once we discretize the space, as we discuss later
in Sect. 2, this issue is no longer relevant.
\end{thebibliography}
\end{document}